\documentclass[aps,ams,prb,groupedaddress,showpacs]{revtex4}

\usepackage{graphicx,amsmath,amssymb,bm,amssymb}

\begin{document}
\def\boldsymbol#1{\mbox{\boldmath$#1$}}

\title{Propagation of coherent waves in elastically scattering media}

\author{Oded Agam}
\affiliation{The Racah Institute of Physics, The Hebrew University,
Jerusalem, 91904, Israel}

\author{A. V. Andreev}

\affiliation{Department of Physics, University of Washington,
  Seattle, Washington, 98195-1560, USA}

\author{B. Spivak}

\affiliation{Department of Physics, University of Washington,
  Seattle, Washington, 98195-1560, USA}

\date{\today}

\begin{abstract}
A general method for calculating statistical properties of speckle
patterns of coherent waves propagating in disordered media is
developed. It allows one to calculate speckle pattern correlations in
space, as well as their sensitivity to external parameters.  
This method, which is similar to the Boltzmann-Langevin approach 
for the calculation of classical fluctuations, applies for a wide 
range of systems: From cases where the ray propagation is diffusive 
to the regime where the rays experience only small angle scattering. 
The latter case comprises the regime of directed waves where rays 
propagate ballistically in space while their directions diffuse. We 
demonstrate the applicability of the method by calculating the
correlation function of the wave intensity and its sensitivity to the 
wave frequency and the angle of incidence of the incoming wave.  
\end{abstract}


\maketitle

\section{Introduction}

Characterization of statistical properties of coherent waves
propagating through an elastically scattering disordered medium is
relevant for a variety of physical situations, ranging from
propagation of electromagnetic waves through interstellar space or
the atmosphere, seismology, and medical imaging by ultrasound or
light, to electron transport in disordered conductors. When coherent
waves propagate through such media their intensity exhibits random,
sample-specific, fluctuations known as speckles. These fluctuations
result from the interference of rays traveling along different
paths. In this article we study the statistics of speckles.

The problem can be characterized by several length scales: The
propagation distance of the ray through the medium, $Z$, the elastic
mean free path, $\ell$, which is the typical distance the ray
travels between two scattering events, and the transport mean free path,
$\ell_{tr}$ which characterizes the typical distance for
backscattering. In the limit of very thin sample, $Z<\ell$, rays
move almost ballistically through the sample, since scattering
porbability is small. This regime has been
extensively studied~\cite{Goodman}. In the opposite limit of a very
wide sample, $Z \gg \ell_{tr}$, the rays propagate diffusively in
the system. This regime has been considered in
Refs.~\onlinecite{ZyuzinSpivak,KaneLee,ZyuzinSpivakRev}. At spatial
scales exceeding the transport mean free path the statistical
properties of speckles in the diffusive regime (excluding
features associated with rare events) are characterized by the
diffusion coefficient and are independent of the details of the
disorder. The crossover between the ballistic and the diffusive
regimes depends, in general, on the features of the disorder. However,
when the typical deflection angle for a single scattering is small, and
therefore the transport mean free path $\ell_{tr}$ is much larger
than the mean free path $\ell$, a third regime emerges. This regime,
known as the directed waves regime, is realized when the sample
width is much smaller than the transport mean free path while it is
much larger than the elastic mean free path, $\ell_{tr}\gg Z\gg
\ell$. In this case, the rays  experience many small angle
scattering events which result in a diffusive dynamics of the ray
direction. The total change in propagation direction, however,
remains small.

\begin{figure}
\includegraphics[width=8.0cm]{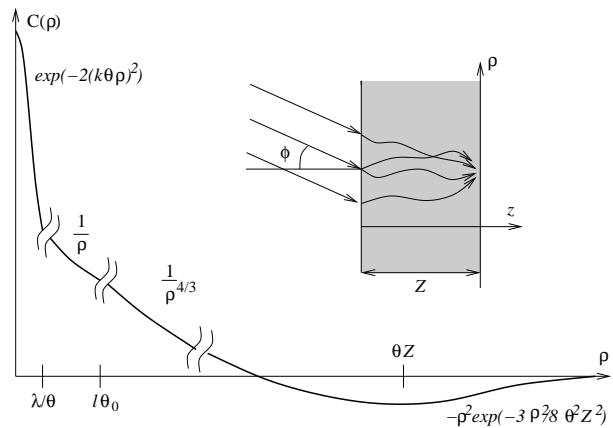}
\caption{The asymptotic behavior of the intensity correlation
  function, ${\cal C}(\rho)$, in the directed waves regime.
$\rho$ is the distance between  the observation points,
 $\lambda$ is the light wavelength, $\ell$ is the elastic mean free path,
$Z$ is the slab width, and $\theta_0$ and $\theta$ are the typical
scattering angle of a ray traveling a distance $\ell$ and $Z$
respectively.} \label{fig:main}
\end{figure}

The focus of our study is on directed waves which are important for
many applications ranging from laser communications in atmosphere to
propagation of acoustic or electromagnetic waves through biological
tissues. Similarly to the  ballistic and the diffusive regimes, the directed
waves regime has  also been studied in many papers (see for example
Refs.~\onlinecite{Tatarski,Kravtsov, Prokhorov,Dashen} and
references therein). However, our results, in many respects, differ
substantially from those obtained in previous studies. One of the
main differences is the slow power law decay of the intensity
correlation function in space, and the change of its sign, see
Fig.~\ref{fig:main}. This
difference affects interpretation of any wave intensity measurement
which uses a finite aperture apparatus.

In this article we develop a general method for calculating speckle
correlations over distances larger than the light wavelength, $\lambda$. This
method, which is similar (but not identical) to the Langevin scheme
for the description of classical fluctuations
\cite{LandauLifshitz,ShulmanKogan,KoganBook}, enables one to treat
both the diffusive and the directed wave regimes on equal footing.
We apply the method to the case of directed waves to evaluate
speckle correlations and their sensitivity to various perturbations,
such as a change in the frequency of the wave, a variation of the
incidence angle, or a change of the refraction index. A short
version of these results was published in
Ref.~\onlinecite{Agam2006}.

The paper is organized as follows. In section \ref{sec:general} we
present the general method describing speckle statistics. In
sections \ref{sec:angular} and \ref{sec:diffusion} we consider its
limiting cases for angular and spatial diffusion. The treatment of
sensitivity of speckle patterns to changes in external parameters is
presented in section \ref{sec:sensitivities}. In section
\ref{sec:directed} we apply our formalism to study speckle
correlations in the directed wave regime, and spatial diffusion. 
Finally, in section \ref{sec:conclusions} we present 
our conclusions. The derivation of the formalism is deferred to the Appendices.

\section{Methods of description  of speckle statistics }
\label{sec:methods}

A paradigm model for propagation of coherent waves through
disordered media is the stationary wave equation for a  scalar field
$\psi({\bf r})$,
\begin{equation}
k^2 n^2({\bf r}) \psi({\bf r}) + \nabla^{2}\psi({\bf r})=0,
\label{eq:St-wave}
\end{equation}
where $k=2\pi/\lambda$ is the wave number, and $n({\bf r})=1+\delta
n({\bf r})$ is the index of refraction. For simplicity we assume
$\delta n({\bf r})$ to be a random Gaussian quantity characterized
by zero average, and isotropic correlation function
\begin{equation}
\langle \delta n({\bf r})\delta n({\bf r'}) \rangle = g(|{\bf
r-r'}|). \label{eq:RefractionCorrelator}
\end{equation}
Here the angular brackets $\langle \ldots \rangle$ denote averaging
over the random realizations of $n({\bf r})$. We assume that the
isotropic function $g(r)$ is characterized by a single correlation length,
$\xi=[\int d^3r r^2 g(r) /3 \int d^3r g(r)]^{1/2}$.

The above model is studied below. The central object of our approach
is the ray distribution function,
\begin{equation}
f({\bf r},{\bf s})\!= \!\int\!\! \frac{p^{2}dp}{2\pi^2} \int\! \!d
\boldsymbol{r}' \psi\left({\bf r}\!-\!\frac{{\bf r'}}{2}\right)
\psi^*\left( {\bf r} \!+\!\frac{{\bf r'}}{2}\right)e^{i p{\bf
s}\cdot {\bf r'}}, \label{Eq:raydis}
\end{equation}
which may be viewed as the density of rays at the point ${\bf r}$
and time $t$ propagating in the direction  specified by the unit
vector ${\bf s}$. In particular the intensity of the wave at the
point ${\bf r}$ is $I({\bf r})\equiv |\psi({\bf r})|^2= \int d^2 s
f({\bf r,s})$.

The ray distribution function $f({\bf r,s})$ is a random, sample
specific quantity whose statistics can be characterized by its
moments. We focus on the first $\langle f({\bf r,s}) \rangle$ and
second $\langle f({\bf r,s})f({\bf r',s'}) \rangle$ moments of this
quantity. These moments quantify the main features of speckle
patterns.

\subsection{General approach to speckle statistics}
\label{sec:general}

In this subsection we discuss a general approach to describe
speckles of coherent waves that is valid both in the ballistic and
diffusive regimes, and holds for a general angular dependence of the
scattering amplitude at a single scatterer.

A general method for calculating moments of the ray distribution
function is the disorder diagram technique~\cite{Abrikosov}. If
$\ell \gg \lambda$, and on the length scale $|{\bf r-r'}|>\lambda$,
this formalism can be reduced to a set of equations for the average
distribution function, $\langle f({\bf r,s}) \rangle$, and the
correlation function of the ray distribution function fluctuations, 
$\langle \delta f({\bf r,s})\delta f({\bf r',s'})
\rangle$, where $\delta f- f-\langle f \rangle$. These equations
describe speckles on various length scales: From the ballistic
regime to the diffusive limit, and are similar, but not identical,
to the Boltzmann-Langevin equations in the kinetic theory of
classical
particles\cite{ShulmanKogan,KoganBook,GurevichGanzevichKatilus}. Thus $\langle f({\bf r,s}) \rangle$ and
$\langle \delta f({\bf r,s})\delta f({\bf r',s'}) \rangle$ can be
deduced from the following set of equations:
\begin{equation}
 {\bf s}\cdot \frac{\partial \langle
f({\bf r,s}) \rangle}{\partial
{\bf r}}=I_{st}[\langle f({\bf r,s})\rangle] \equiv \int d^{2}s' W({\bf s-s'})
 \left( \langle f({\bf r}, {\bf s}')\rangle-  \langle f({\bf r},
{\bf s})\rangle \right), \label{av-ray-dist}
\end{equation}
\begin{equation}
{\bf s}\cdot \frac{\partial \delta f({\bf r,s})}{\partial {\bf r}}-
I_{st}\{\delta f({\bf r,s}) \}= {\cal L}({\bf r}, {\bf s}),
\label{eq:langevin}
\end{equation}
where the integral over the ray directions, ${\bf s}$, is
normalized to unity, $\int d^2 s =1$, and the Langevin sources,
${\cal L}({\bf r}, {\bf s})$, have zero mean and correlations of the
form:
\begin{equation}
\langle {\cal L}({\bf r},{\bf s}) {\cal L}({\bf r'}, {\bf
s'})\rangle = \frac{2\pi }{k^2}\,\delta({\bf r}\!-{\bf r'})
 \left[ \delta({\bf s}\!-\!{\bf s'})\langle
f({\bf r,s})\rangle \int d^{2} \tilde{s} W({\bf s}\!-\! \tilde{\bf s})
\langle f ({\bf r}, \tilde{\bf s})\rangle - \langle f({\bf
r,s})\rangle W({\bf s}\!-\!{\bf s'}) \langle f ({\bf r},{\bf
s'})\rangle \right].
 \label{eq:langevinsources}
\end{equation}
Here $W({\bf s-s'})$  is the probability, per unit length, for
scattering between propagation directions ${\bf s}$ and ${\bf s}'$.
The mean free path $\ell$ and the transport mean free path
$\ell_{tr}$ are expressed in terms of $W({\bf s-s'})$ as
\begin{eqnarray}
    \label{eq:l_def}
    \ell^{-1}&=&
     \int d\boldsymbol{s}' W(\boldsymbol{s}-\boldsymbol{s}'), \quad
     \ell_{tr}^{-1}=\int d\boldsymbol{s}'
     (1-\boldsymbol{s}\cdot
     \boldsymbol{s}')W(\boldsymbol{s}-\boldsymbol{s}').
\end{eqnarray}

In the Born approximation the scattering probability can be
expressed in terms of the refraction index correlator, 
Eq.~(\ref{eq:RefractionCorrelator}), as
\begin{equation}
W({\bf s})=\frac{k^4}{\pi} \int d^{3}r g({\bf r}) e^{i k {\bf s} \cdot
{\bf r}}. \label{kernel-wave}
\end{equation}
The derivation of Eqs.~(\ref{av-ray-dist}-\ref{eq:langevinsources}),
using the standard impurity diagram technique~\cite{Abrikosov}, is
presented in Appendix \ref{sec:Appendix_a}.
On spatial scales larger than $\ell$, and $\ell_{tr}$ it is possible
to simplify Eqs.~(\ref{av-ray-dist}-\ref{eq:langevinsources})
reducing them to a diffusion-type equations. Another simplification
occurs if the scattering angle at a single impurity is small. Then
at lengths greater than the mean free path the change of direction
of the wave propagation is described by diffusion in the angular
space. The simplified form of the general formalism in these two
limits is considered in Sections \ref{sec:angular} and
\ref{sec:diffusion}.

Qualitatively the form of the correlation function of the random
sources, Eq.~(\ref{eq:langevinsources}), can be understood as
follows. Inside the random medium the propagating wave can be viewed
as a random superposition of plane waves arriving from different
directions. The relative phases of the different plane waves are
uncorrelated. Let us consider scattering of this incident wave at a
given impurity. Denoting the amplitude of the wave incident in the
direction $\bm{s}$ by $i(\bm{s})$ we can express the angular
dependence of the the outgoing wave, $o(\bm{s})$, as
\[o(\bm{s})= i(\bm{s}) +2 ik \int d\bm{s}'F(\bm{s}, \bm{s}') i(\bm{s}'),\]
where $F(\bm{s}, \bm{s}')$ is the scattering amplitude. The
intensity of the outgoing wave in the direction $\bm{s}$ is
\begin{equation}\label{eq:outgoing_qualit}
     |o(\bm{s})|^2= |i(\bm{s})|^2- 4k \int d\bm{s}' \mathrm{Im}\left[ F(\bm{s},
\bm{s}')i^*(\bm{s})i(\bm{s}')\right]+ 4k^2 \left| \int
d\bm{s}'F(\bm{s}, \bm{s}') i(\bm{s}')\right|^2 .
\end{equation}
The flux into direction $\bm{s}$ due to scattering, $j(\bm{s})=|o(\bm{s})|^2 -
|i(\bm{s})|^2$, is a random quantity. Since the amplitudes
$i(\bm{s})$ of the incident wave are uncorrelated for different
directions, $\langle i(\bm{s}) i^*(\bm{s}') \rangle \sim \delta
(\bm{s}-\bm{s}') \langle f(\bm{s})\rangle$, the average flux is
given by
\begin{equation}\label{eq:outgoing_average}
    \langle j(\bm{s}) \rangle= - 4k \langle f(s) \rangle \mathrm{Im} [F(\bm{s},
\bm{s})] + 4k^2 \int d\bm{s}' |F(\bm{s}, \bm{s}')|^2 \langle
f(\bm{s}') \rangle =4k^2 \int d\bm{s}' |F(\bm{s}, \bm{s}')|^2
[\langle f(\bm{s}') \rangle- \langle f(\bm{s}) \rangle ] ,
\end{equation}
in agreement with Eq.~(\ref{av-ray-dist}). The last equality in
Eq.~(\ref{eq:outgoing_average}) follows from the optical theorem,
$\mathrm{Im} [F(\bm{s}, \bm{s})]=k \int d\bm{s}' |F(\bm{s},
\bm{s}')|^2 $.

For a specific realization of the incident wave, the flux scattered
in direction $\bm{s}$ differs from its average. In the spirit of the
Boltzmann-Langevin approach one has to evaluate the fluctuations of
microscopic fluxes in the ${\bf s}$ space and substitute them into
the kinetic equation as random sources ${\cal
L}(\bm{s})\sim j(\bm{s})$, see
Eq.~(\ref{eq:langevin}). Thus, for the correlation function of these
quantities, $ \langle{\cal L}(\bm{s}){\cal L}(\bm{s}') \rangle \sim
\langle j(\bm{s})j(\bm{s}') \rangle$,
and using Eq.~(\ref{eq:outgoing_qualit}) one gets the estimate,
\[
\langle{\cal L}(\bm{s}){\cal L}(\bm{s}') \rangle  \sim
 \delta(\bm{s}-\bm{s'}) \langle f(\bm{s}) \rangle \int d
\tilde{\bm{s}}\,|F(\bm{s},\tilde{\bm{s}})|^2 \langle
f(\tilde{\bm{s}}) \rangle - \langle f(\bm{s}) \rangle \langle
f(\bm{s}') \rangle |F(\bm{s},\bm{s}')|^2.
\]
in agreement with Eq.~(\ref{eq:langevinsources}). Here we took into
account the fact that in the limit $\ell\gg \lambda,
F(\bm{s},\bm{s}')$ the main contribution to the flux correlations
comes from the middle term in the right hand side of
Eq.~(\ref{eq:outgoing_qualit}).

\subsubsection{A comparison between
Eqs.~(\ref{av-ray-dist}-\ref{eq:langevinsources}) and
the Langevin description of classical fluctuations. }

It is instructive to compare the method describing classical
kinetics of particles\cite{ShulmanKogan,KoganBook,GurevichGanzevichKatilus}
 with the description of coherent wave
propagating through a disordered media expressed by
Eqs.~(\ref{av-ray-dist}-\ref{eq:langevinsources}). Consider noninteracting
particles propagating in a scattering medium, and let
$\tilde{f}({\bf r,s},t)$ denote their distribution function in phase
space. The scattering process of the particle is random in time and
space. This randomness leads to temporal fluctuations of the
distribution function $\tilde{f}$ even when the incident particle
flux is stationary. It is, therefore, natural to decompose the
distribution function $\tilde{f}({\bf r,s};t)$ into a sum of its
average, $\langle \langle\tilde{f}({\bf r,s};t)\rangle \rangle$, and
fluctuating part, $\delta \tilde{f}({\bf r,s};t)$, characterized by
the correlation function $\langle \langle \delta \tilde{f}\delta
\tilde{f}\rangle \rangle$. Here $\langle\langle \cdots
\rangle\rangle$ denotes the averaging over time, or over the
statistical ensemble~\cite{clav}.

When the elastic mean free path is much larger than the disorder
correlation length, $\ell\gg \xi$, The average distribution function
satisfies the Boltzmann kinetic equation:
\begin{eqnarray}
&& \frac{\partial  \langle \langle \tilde{f}({\bf r,s},t)\rangle
\rangle}{c\,
\partial t}+ {\bf s}\cdot \frac{\partial \langle \langle \tilde{f}({\bf
r,s};t)\rangle \rangle}{\partial {\bf r}}=
I_{st}\{\tilde{f}({\bf r,s};t)\}\equiv \nonumber \\
&&\int d {\bf s}'[I_{st}^{(+)}({\bf s,s'})+I_{st}^{(-)}({\bf
s,s'})]= \int d^{2}{\bf s}'\,
 W({\bf s\!-\!{\bf s}'}) \left( \langle \langle \tilde{f}({\bf r},
{\bf s}';t) \rangle\rangle-  \langle \langle \tilde{f}({\bf r}, {\bf
s};t)\rangle\rangle \right). \label{kinav}
\end{eqnarray}
where $c$ is the particle velocity, $I_{st}^{(+)}({\bf s,s'})$
denotes the particle flux from $\bm{s}'$ to $\bm{s}$ due to
collisions, and $I_{st}^{(-)}({\bf s,s'})$ denotes the particle flux
from $\bm{s}$ to $\bm{s}'$,
\begin{eqnarray}
I_{st}^{(+)}({\bf s,s'})&=&
 W({\bf s\!-\!\tilde{s}}) \tilde{f}({\bf r},{\bf s}';t)
\left( 1 \pm  \tilde{f}({\bf r}, {\bf  s};t) \right), \nonumber
\\
I_{st}^{(-)}({\bf s,s'})&=& -
 W({\bf s-s'})  \tilde{f}({\bf r},{\bf s},t)
\left( 1 \pm  \tilde{f}({\bf r}, {\bf  s'};t) \right). \nonumber
\end{eqnarray}
The $\pm$ sings in front of $\tilde{f}({\bf r}, {\bf s'};t)$
correspond to boson (+) and fermion ($-$) statistics. Notice however
that the quadratic terms in $\langle \langle \tilde{f}({\bf
r,s},t)\rangle\rangle$ cancel out in the Boltzmann equation
(\ref{kinav}) and regardless of the particle statistics.

The statistical behavior of the fluctuations of the distribution function,
$\delta \tilde{f}({\bf  r,s};t)$, may be deduced from the Langevin
equation~\cite{ShulmanKogan},
\begin{equation}
\frac{\partial  \delta \tilde{f}({\bf r,s},t) }{c\, \partial t}+ {\bf s}\cdot
\frac{\partial \delta  \tilde{f}({\bf r,s},t)}{\partial {\bf r}}=
 I_{st}\{\delta \tilde{f}({\bf r,s})\} + \tilde{I}_{L}({\bf r}, t),
\label{Classical-langevin}
\end{equation}
where $\tilde{I}_L$ represents a random Langevin source with vanishing
expectation value and two point correlation function given by
\begin{equation}
\langle \langle \tilde{I}_{L}({\bf r,s},t) \tilde{I}_{L}({\bf r',s'},t') \rangle
\rangle = \delta (t\!-\!t')\delta({\bf  r\!-\!r'})\left\{ \delta({\bf s-s'}) \!
\int ds'' [I_{st}^{(+)}({\bf s,s''})+I_{st}^{(-)}({\bf s,s''})]-
[I_{st}^{(+)}({\bf s,s'})+I_{st}^{(-)}({\bf s,s'})]\right\}.
\label{kinSK}
\end{equation}
The classical limit of this equation corresponds to $\langle\langle
\tilde{f}({\bf r},{\bf s},t) \rangle\rangle \ll 1$. In this case
particle statistics are irrelevant. The description of the evolution
of the average ray distribution function, by the Boltzmann kinetic
equation of a classical particle holds as long as $\ell\gg
\xi,\lambda$. The above formulaes have the following
interpretation\cite{ShulmanKogan,KoganBook}: The scattering processes are
instantaneous and local therefore the correlation function of
Langevin sources (\ref{kinSK}) is proportional to $\delta({\bf
r-r'}) \delta(t-t')$. Thus scattering events generate correlations
of Langevin sources that are nonlocal only in the space of the
particle directions. These are described by the four terms in the
curly brackets. The first two terms, proportional to
$\delta(\bm{s}-\bm{s}')$, describe self-correlation generated by  flux
of particles which scatter from the state $\bm{s}$ to
an arbitrary state $\bm{s}''$ or vice versa.
The two other terms in the curly brackets correspond to scattering
events from ${\bf s}$ to ${\bf s'}$, or back.

The set of Eqs.~(\ref{kinav}-\ref{kinSK}) describing the kinetics of
classical particle and that of
Eqs.~(\ref{av-ray-dist}-\ref{eq:langevinsources}) describing
coherent waves have a similar form. We would like to point out
significant differences originating from the different nature of
fluctuations. A stationary coherent wave propagating through a
disordered sample experiences no temporal fluctuations. In this case
the spatial fluctuations of $f({\bf r,s})$ result from the random
nature of the interference processes associated with different
quasiclassical wave propagation paths. As a result the random
sources, Eq.~(\ref{eq:langevinsources}), are $\delta$-correlated in
space and do not depend on time. In contrast, in the case of
classical particles $\tilde {f}$ fluctuates both in space and in
time, and consequently the random classical sources,
Eqs.~(\ref{Classical-langevin}-\ref{kinSK}) are $\delta$-correlated
both in space and in time.

Another significant difference manifests itself in dramatically
different sensitivities of these two phenomena to small changes of
parameters, such as particle's velocities (or wavelength),
frequencies, and configuration of the scattering potential. In the
case of classical particles the correlators $\langle\langle
\tilde{f}\rangle\rangle$ and $\langle\langle
\delta \tilde{f}\delta \tilde{f}\rangle\rangle$ are insensitive to these changes
as long as the scattering probability $W(\bm{s}-\bm{s}')$ does not
depend on the wave length or the energy of the particles. In
contrast, the coherent speckles exhibit very strong sensitivity to
these changes. As a result the form of the correlation functions of
the random sources describing these sensitivities, see
Eq.~(\ref{eq:langevin-sensitivity}), is very different from that in
Eq.~(\ref{kinSK}).

\subsection{Angular diffusion} \label{sec:angular}

As mentioned above the solutions of
Eqs.~(\ref{av-ray-dist}-\ref{eq:langevinsources}) provide
description of $\langle f({\bf r,s}) \rangle$ and $\delta \langle
f({\bf r,s})\delta f({\bf r',s'}) \rangle$ on the resolution where
$|{\bf r-r'}|>\lambda$. A simplified description is obtained when
the required resolution is over larger length scales. Consider the
case $|{\bf r-r'}|\gg \ell \theta_0$, where $\theta_0=\lambda/\xi$ is
the typical scattering angle over a distance of the order of the
mean free path (notice that the Born approximation implies that $\ell
\theta_0 \gg \lambda$). The reduction of
Eqs.~(\ref{av-ray-dist}-\ref{eq:langevinsources}), for this case, is
similar in spirit to the standard way by which Boltzmann equation is
reduced to the diffusion equation. It follows from the assumption
that $f({\bf s,r})$ changes slowly as function of ${\bf s}$ on the
scale of order $\theta_{0}$. The resulting formulae, provided below,
describe diffusive spreading of the rays in the space of directions,
${\bf s}$. Equation (\ref{av-ray-dist})  reduces to
\begin{equation}
{\bf s} \cdot \frac{\partial \langle f({\bf r,s} \rangle}{\partial {\bf r}}
 = D_\theta \nabla_s^2
 \langle f({\bf r,s}) \rangle, \label{angular-av}
 \end{equation}
where
\begin{equation}
D_\theta=\frac{1}{2}\, \ell_{tr}^{-1}
\end{equation}
is the diffusion constant in the space of angles, ${\bf s}$, and
\begin{equation}
\nabla_{s}= \hat{\theta} \frac{\partial}{\partial \theta}+
\frac{\hat{\phi}}{\sin(\theta)} \frac{\partial}{\partial \phi}
\end{equation}
 is the gradient operator, with the unit vectors
$\hat{\phi}= (-\sin \phi,\cos \phi,0)$
and $\hat{\theta}=(\cos \phi \cos \theta, - \sin \phi\cos
\theta,-\sin \phi)$ (here $\theta$ and $\phi$ are the angles
associated with polar coordinates).

The fluctuations of the ray distribution function in this case are
described by the equation
\begin{eqnarray}
{\bf s} \cdot \frac{\partial \delta f({\bf r,s})}{\partial {\bf r}}
&=&\nabla_s \left[
 D_\theta  \nabla_s
 \delta f({\bf r,s}) -{\bf j}^{L}({\bf r},{\bf s})\right], \\
 \label{eq:langevin_angle_3}
\end{eqnarray}
where the Langevin current sources, ${\bf j}^{L}({\bf r},{\bf s})$,
are correlated as
\begin{eqnarray}
\langle j^L_{\alpha} ({\bf r},{\bf s}) j^L_{\beta}(\tilde{\bf
r},\tilde{\bf s})\rangle &=&\frac{2 \pi D_{\theta}\langle f\rangle
^{2}}{k^{2}}\delta_{\alpha\beta} \delta ({\bf s}\!-\!\tilde{\bf s})
\delta ({\bf r}\!-\! \tilde{\bf r}),
\label{eq:langevin_angle}
\end{eqnarray}
where the indices $\alpha$ and $\beta$ denote the vector components in
the two-dimensional space of directions that is tangential to the unit 
sphere $|s|=1$.

\subsection{Diffusion in real space} \label{sec:diffusion}

If one is concerned with an even cruder resolution, where $|{\bf
r-r'}|\gg \ell_{tr}$, the effective description of the system
employs the diffusion equation in real space. In this case $f({\bf
r,s})$ is assumed to be a nearly isotropic function of ${\bf s}$,
and a slow function of $\bm{r}$.  Then
Eqs.~(\ref{av-ray-dist}-\ref{eq:langevinsources}) can be reduced to
the following set of Diffusion-Langevin
equations~\cite{ZyuzinSpivak,ZyuzinSpivakRev}. Namely, expressing
the wave intensity, $I(\bm{r})$, at point $\bm{r}$ as $I({\bf r})=
\int d^2s f(\bm{r},{\bf s})$, one can reduce Eq.~(\ref{av-ray-dist})
to the Laplace equation,
\begin{eqnarray}
\nabla^2 \langle I({\bf r})\rangle =0, \label{av-diff}
\end{eqnarray}
while the correlator of the intensity fluctuations, $\delta I=
I-\langle I \rangle$,  can be deduced
from the flux conservation condition,
\begin{eqnarray}
\nabla  \cdot\delta {\bf J} =0, \label{eq:incompress}
\end{eqnarray}
with
\begin{eqnarray}
\delta {\bf J}= - D \nabla \delta I +{\bf J }^{L}.  \label{eq:J}
\end{eqnarray}
Here $D=\ell_{tr}/3$  is the diffusion constant in real space
(notice that according to our convention the diffusion constant has
dimensions of length). The Langevin current sources, ${\bf J}^{L}$,
have a vanishing expectation value and are characterized by the
correlation function:
\begin{eqnarray}
\langle J_{\alpha}^L({\bf r}) J_{\beta}^L({\bf r'})\rangle=
\frac{\lambda^2 D}{2\pi}
 \langle I ({\bf r}) \rangle^{2} \delta_{\alpha \beta}  \label{eq:JLangevin}
\delta ({\bf r}-{\bf r'}).
\end{eqnarray}
The boundary conditions for these equations are the conventional
conditions for the diffusion equation: $\delta I=0$ at open
boundaries, and ${\bf J}\cdot \bm{n}=0$, with $\bm{n}$ being the
normal to the boundary, at closed boundaries.

\subsection{Sensitivities of speckles to changes of external parameters.}
\label{sec:sensitivities}

The interfering waves travel along different paths, and the lengths
of these paths are much longer than the wave length. Therefore the
phases accumulated along each path are very sensitive to changes of
external parameters such as the wave number $k$, the incidence angle
of the incoming wave, or a smooth change in the refractive index
$\Delta n({\bf r})$. We will characterize these changes by the
control parameter $\gamma({\bf r})= \Delta k+ k \Delta n({\bf
r})$ where $\Delta k$ denotes a change in the wave number, $k$.
The formalism presented above may be straightforwardly generalized to
calculate the sensitivity of the speckle pattern to various external
perturbations. The sensitivity of the speckle pattern
can be characterized by the correlator of the ray
distribution functions at different values of the control parameter,
$\langle \delta f({\bf r,s},0)\delta f({\bf r,s},\gamma) \rangle $.
In order to evaluate it  Eq.~(\ref{eq:langevin}) should be replaced
by two equations. One for $\delta f({\bf r,s},0)$, and another for
$\delta f({\bf r,s},\gamma)$. The form of these equations is
precisely that of Eq.~(\ref{eq:langevin}), however the Langevin
sources now depend on the perturbation parameter $\gamma$. Namely
\begin{equation}
{\bf s}\cdot \frac{\partial \delta f({\bf r,s};\gamma)}{\partial {\bf r}}-
I_{st}\{\delta f({\bf r,s};\gamma) \}= {\cal L}({\bf r}, {\bf s};\gamma), \label{eq:langevin-sen}
\end{equation}
where ${\cal L}({\bf r},{\bf s};\gamma)$ denotes the Langevin source
associated with the value $\gamma$ of the perturbation. The average
of the Langevin sources vanishes. Their correlation function, at
different points in space and different values of the control
parameter, is given by
\begin{widetext}
\begin{equation}
\langle {\cal L}({\bf r},{\bf s};0) {\cal L}({\bf r'}, {\bf
s'};\gamma )\rangle \!=\! \frac{\pi}{k^2}\delta({\bf r}\!-{\bf r'})
\sum_{\nu=\pm} \left[ \delta({\bf s}\!-\!{\bf s'}) f_\nu({\bf r,s})
\int d^{2}{\bf s}_{1} W({\bf s}\!-\!{\bf s}_{1}) f_{-\nu}({\bf
r},{\bf s}_{1}) - f_\nu({\bf r},{\bf s}) W({\bf s}\!-\!{\bf s'})
f_{-\nu}({\bf r},{\bf s'}) \right], \label{eq:langevin-sensitivity}
\end{equation}
\end{widetext}
where $f_\pm ({\bf r,s})$ satisfies the equation,
\begin{equation}
{\bf s} \cdot \frac{\partial  f_\pm({\bf r,s})} {\partial {\bf r}}-
I_{st}\{f_{\pm }({\bf r,s})\}=\pm i \gamma  f_\pm({\bf r,s})
.\label{eq:kinetic-senstitivity}
\end{equation}
At free boundaries, the boundary conditions for the functions
$f_\pm({\bf r,s})$ coincide with the standard boundary conditions
for the Boltzmann equation. At the boundary with an incident
radiation, denoted by ${\cal S}$, the functions $f_\pm({\bf r,s})$
are determined by the parametric correlations in the incident wave,
i.e.
\begin{equation}
\left. f_+({\bf r},{\bf s})\right|_{\bm{r} \in {\cal S}}\!=
\!\int\!\! \frac{p^{2}dp}{2\pi^2} \int\! \!d \boldsymbol{r}'
\psi_\gamma\left({\bf r}\!-\!\frac{{\bf r'}}{2}\right)
\psi_0^*\left( {\bf r} \!+\!\frac{{\bf r'}}{2}\right)e^{i p{\bf
s}\cdot {\bf r'}}. \label{eq:raydis_parametric}
\end{equation}
Here the subscript of the wave amplitude $\psi$ denotes the value of
the parameter $\gamma$. The corresponding equation for $f_-({\bf
r},{\bf s})$ is obtained from Eq.~(\ref{eq:raydis_parametric}) by
interchanging the subscripts: $\gamma \leftrightarrow 0$.

When the external perturbation is associated with a change in the
incidence angle of the incoming wave,
Eq.~(\ref{eq:langevin-sensitivity}) still holds, however, both $f_+
({\bf r,s})$ and  $f_- ({\bf r,s})$ satisfy the same equation
(\ref{angular-av}). The difference between $f_+ ({\bf r,s})$ and
$f_- ({\bf r,s})$ arise from the boundary
conditions,(\ref{eq:raydis_parametric}). We shall elaborate on this
issue in Section \ref{sec:directed-senstivity}.

The above formulae describe the speckle sensitivity
on the resolution scale larger than the wavelength. As discussed in
the previous section the formalism simplifies for lower resolution.
 We conclude this section by providing the relevant formulas for the
case of angular diffusion, and diffusion is real space.

\subsubsection{Sensitivity in the case of angle diffusion}
\label{sec:angle-sensitivity}

If the typical scattering angle at a single impurity is small and
wave propagation length exceeds the mean free path, $\ell$, the
equation for the fluctuations in the ray distribution function is
\begin{eqnarray}
{\bf s} \cdot \frac{\partial \delta f({\bf r,s};\gamma)}{\partial {\bf r}}
&=&\nabla_s \left[
 D_\theta  \nabla_s
 \delta f({\bf r,s};\gamma) -{\bf j}^{L}({\bf r},{\bf s};\gamma)\right],
 \label{eq:langevin_angle_sen1}
\end{eqnarray}
where the Langevin current sources, ${\bf j}^{L}({\bf r},{\bf
  s};\gamma)$ depend on the perturbation $\gamma$. These have
  zero mean and correlation function given by
\begin{eqnarray}
\langle j^L_{\alpha} ({\bf r},{\bf s};0) j^L_{\beta}(\tilde{\bf
r},\tilde{\bf s};\gamma)\rangle &=&\frac{2 \pi D_{\theta} f_+({\bf
r},{\bf s}) f_-({\bf r},{\bf s})}{k^{2}}\delta_{\alpha\beta} \delta
({\bf s}\!-\!\tilde{\bf s}) \delta ({\bf r}\!-\! \tilde{\bf r}),
\label{eq:langevin_angle_sen}
\end{eqnarray}
where $f_\pm({\bf r},{\bf s})$ satisfy the equation
\begin{equation}
 {\bf s} \cdot \frac{\partial f_\pm({\bf r,s})}{\partial {\bf r}}
= D_\theta \nabla_s^2  f_\pm({\bf r,s})\pm i \gamma f_\pm({\bf r,s}). \label{angular-av-sen}
 \end{equation}

\subsubsection{Sensitivity in the case of the real space diffusion}
\label{sec:diffusion-sensitivity}

Finally, on spatial scale larger than the transport mean free path
$\ell_{tr}$, the sensitivity of the speckle  pattern may be
described by the current conservation condition,
\begin{eqnarray}
\nabla \cdot \delta {\bf J}=\nabla \cdot\left(- D \nabla \delta I +{\bf
    J}^{L} \right)=0,  \label{sendif1}
\end{eqnarray}
where the Langevin current sources, at different values of the
perturbation parameter, $\gamma$, are correlated as
\begin{eqnarray}
\langle J_{\alpha}^L({\bf r};0) J_{\beta}^L({\bf r'};\gamma)\rangle=
\frac{\lambda^2 D}{2\pi}
  I_+ ({\bf r}) I_- ({\bf r}) \delta_{\alpha \beta}
\delta ({\bf r}-{\bf r'}), \label{sendif2}
\end{eqnarray}
and $I_\pm({\bf r})$ satisfies the equation
\begin{eqnarray}
D \nabla^2 I_\pm ({\bf r}) \pm i \gamma I_\pm ({\bf r})
=0.  \label{sendif3}
\end{eqnarray}

\section{Evaluation of speckle correlation functions and speckle sensitivities
to changes of external parameters}
\label{sec:directed}

In this section we shall illustrate the use of the formalism
developed in the previous section. To this end we will consider the
correlation function of speckles and their sensitivity to various
perturbations in the regimes of directed waves as well as for
diffusion in real space.

\subsection{Speckles in the regime of directed waves}

Consider a situation in which a  wave of intensity $I_0$ is incident
on a disordered slab of thickness $Z$, as shown in the inset of
Fig.~\ref{fig:main}. The slab thickness is assumed to be much
smaller than the transport mean free path and much larger than the
elastic mean free path, $\ell_{tr} \gg Z \gg \ell$. Thus rays
diffuse in angle, but their total change of direction is small. In
this regime of directed waves it will be convenient to choose the
coordinate system ${\bf r}=(\boldsymbol{\rho}, z)$ where $z$ is the
the direction of the wave propagation in the absence of disorder
($\delta n({\bf r})=0$), and $\boldsymbol{\rho}$ denotes a two
dimensional vector in the plane perpendicular to the $z$-axis.
Similarly we decompose the vector of the ray direction as ${\bf
s}=({\bf
  s}_\perp),s_z$, where $s_z$ denotes the component in the $z$ direction,
while ${\bf s}_\perp$ is a two dimensional vector in the
perpendicular plane. The rays of directed waves are almost parallel
to the $z$ axis and therefore $s_z \approx 1$, i.e.  ${\bf s}\approx
({\bf s}_\perp,1)$. If we denote by $\theta$ the typical ray angle
at $z=Z$, then the latter approximation holds as long as $\theta \ll
1$. The results which we present below are calculated to leading
order in the small parameter $\theta$.

It is instructive to start with understanding the classical
evolution of the average ray distribution function in the regime of
directed waves. For this purpose we solve Eq.~(\ref{angular-av}) for
the case where a single ray moving in the $z$ direction, impinges
upon the slab at the origin ${\bf r}=0$. The assumption that ${\bf
s} \simeq ({\bf s}_\perp,1)$ allows one to reduce
Eq.~(\ref{angular-av}) to
\begin{equation}
\frac{ \partial \langle f({\bf r,s})\rangle}{\partial z}+ {\bf s}_\perp
\cdot \frac{\partial  \langle f({\bf r,s})\rangle }{\partial
  \boldsymbol{\rho}} - D_\theta \frac{ \partial^2
 \langle f({\bf r,s}) \rangle}{\partial {\bf s}_\perp^2}=0. \label{para-av}
 \end{equation}
The boundary conditions are
\begin{eqnarray}
\left.\langle f({\bf r},{\bf s}) \rangle \right|_{z=0}= i_0 \delta
(\boldsymbol{\rho}) \delta ({\bf s}_\perp),
\end{eqnarray}
where the amplitude $i_0$ denotes the incident ray intensity. The
solution of the above problem takes the form
\begin{eqnarray}
\langle f({\bf r},{\bf s}) \rangle = \frac{3 i_0}{4 \pi^2 D_\theta^2
  z^4} \exp \left[-\frac{ 3 \boldsymbol{\rho}^2}{D_\theta z^3}+ \frac{ 3
  {\bf s}_\perp  \boldsymbol{\rho}}{D_\theta z^2}- \frac{{\bf
  s}_\perp^2}{D_\theta z} \right]. \label{AvRayDis}
\end{eqnarray}
It demonstrates the diffusive behavior of the ray direction as it
propagates in the slab, $|{\bf s}_\perp|^2 \sim D_\theta z$. It also
shows that deviations in real space grow in a superdiffusive
manner\cite{Jayannavar82}, $\rho^2 \sim D_\theta z^3$.

After this preliminary consideration we turn to study intensity
correlations of directed waves. To be specific we consider a plane
wave (not restricted by a finite aperture) incident on the
disordered slab in the $z$-direction. In this case the average ray
distribution function is independent of the perpendicular coordinate
$\bm{\rho}$ and can be easily obtained by integrating
Eq.~(\ref{AvRayDis}) over $\bm{\rho}$,
\begin{eqnarray}
\langle f(z,{\bf s}) \rangle = \frac{i_0}{4 \pi D_\theta
  z} \exp \left[- \frac{{\bf
  s}_\perp^2}{4 D_\theta z} \right]. \label{eq:f_s,z}
\end{eqnarray}

The intensity correlation function,
\begin{eqnarray}
{\cal C}(\delta {\bf r})\equiv
\langle \delta I({\bf r}) \delta I({\bf r}+\delta \bm{r}) \rangle,\label{def:IntensityCo}
\end{eqnarray}
where $\delta I({\bf r}=I({\bf r})-\langle I({\bf r}\rangle$,
is independent of the transverse coordinate and depends only on the
propagation distance $Z$ and the difference coordinate $\delta
\bm{r}$. The behavior of this correlator as a function of $\delta
\bm{r}=(\bm{\rho},\delta z)$ is strongly anisotropic. Consider first
the case where the observation points are located along the $z$ axis
(i.e. $\bm{\rho}=\bf{0}$) near the point $z=Z$. In
this case we obtain
\begin{equation}
{\cal C}(\delta z)=\frac{I_0^2}{4k^2\theta^4 \delta z^2}, \label{eq:C}
\end{equation}
where $\theta =\sqrt{D_\theta Z} $ is the accumulated scattering
angle, and the condition $\ell \ll \delta z\ll Z$ is assumed. This
formula, which also approximates the behavior for nonzero $\rho$ as
long as $\delta z \gg \rho/\theta$, matches the results for the
diffusive case \cite{ZyuzinSpivak,ZyuzinSpivakRev}, $Z\gg \ell_{tr}$
when $\theta$ is of order unity.

A more complex behavior of the correlation function appears  when
$\delta z<\rho/\theta$, i.e. when  the observation points are
located essentially in the plane perpendicular to the $z$-axis.  A
general formula for ${\cal C}(\boldsymbol{\rho})$, in this case, is
derived in Appendix B. The expression takes the form
\begin{eqnarray}
{\cal C}(\rho)&=&\frac{I_0^2}{4 D_\theta k^2}
\int_0^{Z-\ell}\!\frac{d\zeta}{\zeta-Z}\int_0^\infty \!\! dq q J_0(q
\rho) \frac{d}{d\zeta} \exp \left[-\frac{2}\ell \int_0^\zeta
d\eta \left\{ 1- \tilde{g}\left(\frac{q}{k}\eta\right)\right\}
\right], \label{eq:main}
\end{eqnarray}
where $\tilde{g}(\rho)=\int dz g(\sqrt{\rho^{2}+z^{2}})/\int dz
g(z)$, and $J_0(x)$ is the Bessel function of zeroth order.

The integral in Eq.~(\ref{eq:main}) contains a term
proportional to a $\delta$-function, $\frac{\pi I_0^2}{2 D_\theta
 k^2 Z}\delta(\boldsymbol{\rho})$. This term represents the rapidly
decaying (at $\rho \sim \lambda/\theta$) part of the correlator. It
results from the semiclassical approximation employed in the
derivation of Eqs.~(\ref{av-ray-dist}-\ref{eq:langevinsources}),
which limits the spacial resolution to $\rho > \lambda$. In order to
resolve the spatial structure on smaller scales some of the diagrams
discussed in Appendix \ref{sec:Appendix_a} should be calculated more
accurately. The result of this calculation shows that the $\delta$
function contribution to the correlator ${\cal C}(\rho)$ is in fact
a contribution of the form  $I_0^2 e^{-2(k \theta \rho )^2}$ where
$\theta^2=D_\theta Z$.

As we show below, ${\cal C}(\rho)$ contains also a slowly decaying
term. The latter, which has been overlooked in previous studies,
clearly has important consequences. In order to understand this
term it will be instructive to explain, first, the origin of the
short ranged contribution to  ${\cal C}(\rho)$. As we show now, it
arises from a superposition of statistically independent
contributions of waves moving in all possible directions. Let us
assume that wave function at a given point on the screen is a sum of
plane waves. The distribution of directions of these plane waves is
dictated by the diffusive nature of the rays in the system, thus
\begin{eqnarray}
\psi(\boldsymbol{\rho})= \sum_\nu A_\nu e^{ ik {\bf s}_{\perp,\nu} \cdot
  \boldsymbol{\rho}} \label{stat-psi}
\end{eqnarray}
where ${\bf s}_{\perp,\nu}$ denotes the direction of the $\nu$-th
contribution and $A_\nu$ is the corresponding amplitude. We shall
assume that  $A_\nu$ are statistically independent variables, with
zero mean and fluctuation strength given by
\begin{eqnarray}
 \langle |A_\nu|^2\rangle= \frac{I_0}{4 \pi D_\theta Z}e^{\frac{ |{\bf
 s}_{\perp,\nu}|^2}{4 D_\theta Z}}. \label{amplitude-dis}
\end{eqnarray}
The average $\langle |A_\nu|^2\rangle$ may be interpreted as
the ``classical'' probability to find a plain wave moving in
direction ${\bf s}_{\perp,\nu}$. It may be  obtained from the
solution of Eq. (\ref{para-av}) with boundary conditions which
correspond to an impinging plane wave of density $I_0$,
$\left.\langle f({\bf r},{\bf s}) \rangle \right|_{z=0}= I_0 \delta
({\bf s}_\perp)$.

The above assumptions imply that, at a given point in space,
$\psi(\boldsymbol{\rho})$ is approximately a Gaussian random
variable, as a result of the central limit theorem. Moreover, the wave
function at two different points, $\psi(\boldsymbol{\rho})$ and
$\psi(\boldsymbol{\rho}')$, are also described by a joint Gaussian
distribution function provided the distance between these points is
sufficiently small such that one may assume that the same set of
wavelets arrive to both points.

Assuming the observation points $\boldsymbol{\rho}$ and
$\boldsymbol{\rho}'$ to be sufficiently close to each other,
consider the ensemble average $\langle I(\boldsymbol{\rho})
I(\boldsymbol{\rho}')\rangle = \langle
\psi(\boldsymbol{\rho})\psi^*(\boldsymbol{\rho})
\psi(\boldsymbol{\rho}') \psi^*(\boldsymbol{\rho}') \rangle$. Using
the fact that within a small vicinity of space,
$\psi^*(\boldsymbol{\rho})$ may be considered as a random Gaussian
function, one deduces that $\langle
\psi(\boldsymbol{\rho})\psi^*(\boldsymbol{\rho})\rangle \langle
\psi(\boldsymbol{\rho}') \psi^*(\boldsymbol{\rho}') \rangle+ \langle
\psi(\boldsymbol{\rho})\psi^*(\boldsymbol{\rho}')\rangle \langle
\psi(\boldsymbol{\rho}') \psi^*(\boldsymbol{\rho}) \rangle$, and
hence the density correlation function is given by
\begin{eqnarray}
{\cal C}(\boldsymbol{\rho}-\boldsymbol{\rho}')= \left|\langle
\psi(\boldsymbol{\rho})\psi^*(\boldsymbol{\rho}')\rangle \right|^2
\label{C-stat}
\end{eqnarray}
Now from (\ref{stat-psi}) and the statistical independence of the
amplitudes $A_\nu$ we see that
\begin{eqnarray}
\langle
\psi(\boldsymbol{\rho})\psi^*(\boldsymbol{\rho}')\rangle =
\left\langle \sum_{\nu,\nu'} A_\nu A^*_{\nu '} e^{ik \left( {\bf s}_{\perp,\nu}
\boldsymbol{\rho}-{\bf s}_{\perp,\nu'} \boldsymbol{\rho}')
\right)}\right\rangle =
\sum_\nu  \langle |A_\nu|^2\rangle  e^{ik {\bf s}_{\perp,\nu} \left(
\boldsymbol{\rho}- \boldsymbol{\rho}'\right) }. \label{diagonal}
\end{eqnarray}
The replacement of the above double sum by a sum over one index is
equivalent to the assumption that the interference terms of
different amplitudes average out to zero. This traditional procedure
in semiclassical analysis, known as  the ``diagonal approximation''
leaves only the classical contribution. Thus substituting
(\ref{amplitude-dis}) in (\ref{diagonal}) and replacing the sum over
$\nu$ by an integral over ${\bf s}_\perp$ we obtain an expression
for $\langle
\psi(\boldsymbol{\rho})\psi^*(\boldsymbol{\rho}')\rangle$, and from
(\ref{C-stat}) we conclude that
\begin{eqnarray}
{\cal C}(\boldsymbol{\rho}-\boldsymbol{\rho}')\approx I_0^2 e^{-2(k
\theta |\boldsymbol{\rho}-\boldsymbol{\rho'}| )^2}.
\end{eqnarray}

This expression, which shows very fast decay of correlations on a
scale of order $\lambda/\theta$, has rather limited range of
applicability.  The reason is that the description of the wave
function in the sample, using the superposition of independent plane
waves (\ref{stat-psi}), gives reasonable approximation only 
when the observation points are very close. At larger distances 
diffraction and quantum 
impurity scattering give rise to correlation of rays which manifest
themselves in a slow decay of the intensity correlations, as well 
a change of sign. These effects are described by Eq.~\ref{eq:main},
and illustrated in Fig.~1. At various spatial separations 
$\bm{\rho}$ one can obtain the following asymptotic expressions for 
the intensity correlator:
\begin{equation}
\frac{{\cal C}(\rho)}{I_0^2}\approx \left \{
\begin{array}{ll} e^{-2(k \theta \rho )^2} &
\textrm{if~~~ $\rho \sim\alpha
 \lambda /\theta$}, \\
\frac{b_1}{k^2 \theta^2\ell \theta_0 \rho} &   \textrm{if~~~ $\alpha
\lambda / \theta \ll \rho\ll \ell \theta_0$}, \\
 \frac{ b_2 D_\theta^{2/3}}{k^{2}
\theta^4 \rho^{4/3}} & \textrm{if~~~ $ \ell \theta_0 \ll \rho \ll
\theta Z $} ,\\
\frac{-b_3 \rho^2}{k^2 \theta^6 Z^4}e^{-\frac{3 \rho^{2}}{8
  \theta^2 Z^2}} & \textrm{if~~~ $\theta Z \ll \rho \ll
 \frac{Z\theta^2}{\theta_0}$},
\end{array} \right.
\label{eq:asymptotic}
\end{equation}
where $\alpha^2=\log(k \ell \theta^3/\theta_0)$, $b_1=\int_0^\infty
dx \tilde{g}(x)/\xi$ is a constant of order unity, $b_2=3^{1/3}\Gamma
(5/3)/8\approx 0.163$, and $b_3=27/128\approx 0.21$. The qualitative
 form of the function ${\cal C}(\rho)$ is shown in Fig.~\ref{fig:main}.

In order to clarify the connection between ray diffraction and the
slow decay of the density correlations, let us focus on  the regime
$\ell \theta_0 \ll \rho \ll \theta Z$. Consider two points separated
by a distance $\rho$. The correlations of the wave intensity in
 these points emerge from coherent waves which simultaneously arrive
to the two points. These can be generated by diffraction which acts
as a beam splitter, and modeled by the Langevin sources in
Eq.~(\ref{eq:langevin}). Now, the superdiffusion nature of the ray
dynamics in the sample implies that the relevant points where
diffraction takes place  should be located at distance of order
$\Delta z$ from the screen, where $\rho^2=D_\theta \Delta z^3$. The
wave intensity emitted from these diffraction points decays as
$1/\Delta z^2$, and therefore the correlations which they generate
are proportional to $D_\theta^{2/3}/\rho^{4/3}$.

The above crude argument explains the power law decay of ${\cal
C}(\rho)$, in the regime $\ell \theta_0 \ll \rho \ll \theta Z$. Yet
a closer examination of the integrals leading to this results shows
that the contribution from diffraction points (or Langevin sources)
that are closer to the screen than, $\Delta z=(D_\theta
\rho^2)^{1/3}$ generate anti-correlations, while those that are at
larger distances provide positive correlations. This behavior may be
expected since diffraction points located too close to the screen
generate rays which may arrive to either one of the observation
points but not to both of them,  therefore they lead to
anti-correlated behavior. On the other hand, coherent waves
generated by diffraction that took place at distances larger than
$\Delta z$,  get, in general, to both points, and therefore generate
positive correlations.

From this picture, and the finite width of the slab, it follows that
for sufficiently large distance between the observation points, $
\rho^2 \gg D_\theta Z^3=(Z\theta)^2$, diffraction events can generate only
anticorrelations. Thus ${\cal C}(\rho)$ must experience a sign
change in the vicinity of $\rho=\theta Z$.

Finally, we mention that the tail of the correlation function (the
regime $\rho>Z\theta^2/\theta_0$) is also described by
Eq.~(\ref{eq:main}). However, it  depends on the precise form of the
disorder correlation $g(r)$, since this limit is dominated by rare
scattering events.

The power law nature of the density correlations of directed waves
have important consequences regarding the statistics of the signal
measured by sensors with large apertures compared to the wave length.
Let
\begin{eqnarray}
P=\int d^2\rho d^2 s~ {\bf n}\cdot{\bf s} f({\bf r,s}), \label{power}
\end{eqnarray}
 denote the signal measured by the sensor, where ${\bf n}$ is
a unit vector perpendicular to the sensor surface, and
$\boldsymbol{\rho}$ is a two dimensional vector which
parameterizes the sensor surface.  If the sensor aperture is circular,
with radius $R$, and its surface is perpendicular
to the propagation direction, i.e. ${\bf s \cdot n} \sim 1$, then
the integrated power measured by the sensor  may be
approximated by an integral over the wave density
\begin{eqnarray}
P(R) =\int_{|\boldsymbol{\rho}|<R}  d^2\rho  I({\bf r}),
\end{eqnarray}
where ${\bf r}=(\boldsymbol{\rho},z)$. The random fluctuations of
 $I({\bf r})$ imply that $P(R)$ is also a random quantity.
Its average may be expressed as an integral over $\langle I({\bf r})
 \rangle$, while the variance of its fluctuations is given by
\begin{eqnarray}
\langle \left( \delta P(R)\right)^2 \rangle =\int_{|\boldsymbol{\rho}|,
|\boldsymbol{\rho}'|<R} d^2\rho d^2 \rho'  C({\bf r-r'}),
\end{eqnarray}
where $C({\bf r-r'})$ is the density correlation function
(\ref{eq:main}).

Clearly, the fluctuations of $P(R)$ strongly depend on
the slow power law tails of the correlation function as
well as its sign change. The asymptotic behavior of the variance of
these fluctuations, for a circular sensor with aperture radius $R$ is
given by
\begin{equation}
\frac{\langle(\delta P(R))^2\rangle }{I_0^2 \pi R^2}\approx \left \{
\begin{array}{ll}
\frac{\pi}{2k^2 \theta^2}+\frac{b'_1 R}{k^2
\theta^4 \ell \theta_0} ,&  \frac{\alpha
\lambda}{ \theta} \! \ll \! R \! \ll \! \ell \theta_0 ,\\
 \frac{\pi}{2k^2 \theta^2}+\frac{b'_2 (D_\theta R)^{2/3}}{k^2
 \theta^4},
 & \,\ell \theta_0 \! \ll \! R \!\ll\!
\theta Z , \\
b'_3 \frac{Z}{k^2 \theta R}, & \theta Z \! \ll \! R \! \ll \!
 \frac{Z \theta^2}{\theta_0},
\end{array} \right.
\label{eq:power-asymptotic}
\end{equation}
where $b'_1=2b_1\pi/3$, $b'_2=3^{4/3}\Gamma(5/6) \pi/2^{11/3}
\Gamma(7/6)$, and $b'_3=\sqrt{3/2\pi}$.

\subsubsection{Speckle sensitivity to change of the wave frequency}

Consider the sensitivity of the speckle patterns of directed waves
to a change in the wave frequency: $\Delta \omega=c\Delta k$, where
$c$ is the speed of the wave, and $k$ is the wave number. Using
Eqs.~(\ref{eq:langevin-sen}-\ref{eq:kinetic-senstitivity}), with the
appropriate control parameter, $\gamma=\Delta k$, treated on a
perturbative level, one may identify the scale of the change in the
control parameter, where the new speckle pattern essentially lost
its correlations with the initial one (i.e. the speckle pattern at
$\gamma =0$). For the wave frequency perturbation this scale is
found to be
\begin{equation}
\omega^{*}=  \frac{c}{\theta^2 Z}.
 \label{eq:omega*}
\end{equation}

A qualitative explanation of the scale $\omega^*$ is similar to that
given for the sensitivity of the conductance fluctuations
\cite{LeeStone,AltshulerSpivak}. Let us estimate the characteristic
change in the phase of a typical orbit due to the frequency change
$\Delta \omega$: The typical length spread of the orbits, in the
directed waves regime, as follows from their superdiffusive nature,
is of order $\theta^2 Z$. Therefore the phase difference between a
given orbit and the same orbit different frequency, is of order
$\Delta k Z \theta^2$ where $\Delta k=\Delta \omega/c$ is the change
in the wave number. Thus a complete change of the speckle pattern
occurs when the phase, $\Delta \omega Z \theta^2/c$ is of order one,
namely $\Delta \omega \sim c/Z\theta^2\sim \omega^*$, in agreement
with Ref.~\onlinecite{Dashen}.

\subsubsection{Sensitivity of speckles to change of the angle of incidence}
\label{sec:directed-senstivity}

Consider the case where rays propagate through a disordered slab
whose one edge is located at $z=0$. A plane wave, moving in direction
approximately parallel to the $z$ axis, impinges the slab, at $z=0$.
The speckle pattern formed on the second edge of the slab, at $z=Z$,
will be sensitive to the precise angle, $\phi$, of the incoming wave.
The latter takes the form
$\psi=\sqrt{I_0} \exp[i k z\cos\phi + ik \rho\sin\phi ]$.

As mentioned in the previous section, the sensitivity in this case
is characterized by the correlation function
(\ref{eq:langevin-sensitivity}) of the Langevin sources $\langle
{\cal L}({\bf r},{\bf s};0) {\cal L}({\bf r'}, {\bf s'};\phi
)\rangle$, where both $f_+({\bf r,s})$ and $f_-({\bf r,s})$ satisfy
the same equation
\begin{equation}
{\bf s} \cdot \frac{\partial  f_\pm({\bf r,s})} {\partial {\bf r}}-
I_{st}\{f_{\pm }({\bf r,s})\}=0. \label{ang-sen}
\end{equation}
However their boundary conditions are different. They are determined
by the Wigner transforms of a product of the incoming wave parallel
to the $z$ axis, by the complex conjugate of an incoming wave at
angle $\pm \phi$ (evaluated at $z=0$). Thus the boundary conditions
for Eq.~(\ref{ang-sen}) are
\begin{equation}
f_\pm(\vec{\rho}, z=0)=  I_0 e^{\pm i k {\bf s}_\perp
  \vec{\rho}}\delta ({\bf s}-{\bf s}_0),
\end{equation}
where ${\bf s}_0=(\cos \phi,\boldsymbol{s}_\perp)\approx
(1,\boldsymbol{s}_\perp)$ denotes a unit vector in the direction of
the incoming wave, and  $|\boldsymbol{s}_\perp|=\sin \phi\approx
\phi$, assuming $\phi \ll 1$. Solving the above equations one can
identify the characteristic scale for the change in the incidence
angle:
\begin{eqnarray}
\phi^*= \frac{1}{k Z \theta}. \label{eq:phiSen}
\end{eqnarray}
This result has simple interpretation. Consider a given point on the
screen. The wave intensity at this point is determined by the
interference of all the rays which originate at $z=0$ and reach the
same point. The nature of the ray dynamics, in the directed wave
regime, implies that the the original distance between two rays
which reach the same point at the screen is of order of $Z\theta$.
Now if we change the incidence angle by some small amount $\phi \ll
1$, the phase difference between two such rays is of order $k
Z\theta \phi$, where $k$ is the wave number. The interference of
these rays will be completely different when this phase difference
is of order one, i.e. $k Z\theta  \phi^* \sim 1$. From here we
obtain (\ref{eq:phiSen}).

\subsection{Speckle statistics in the diffusive regime, $Z\gg
\ell_{tr}$}

In what follows we complete the picture of speckle statistics by
presenting the well known results of speckle
correlation functions and sensitivities for the
diffusive regime, $Z \gg \ell_{tr}$.  For simplicity we consider
the situation where $\ell= \ell_{tr}$, and set the resolution
scale to be  larger than the wavelength, $\lambda$. Furthermore,
as in the previous section, we shall
consider the infinite slab geometry shown in Fig.~\ref{fig:main},
and assume that
 plane wave, moving in the $z$ direction, impinges the system at $z=0$.

\subsubsection{The intensity correlation function}

Our first step is to solve Eq.~(\ref{av-diff}) for the average
intensity. The boundary conditions in this case are $I({\bf
r})|_{z=Z}=0$ and $d I({\bf r})/dz|_{z=0}=-J_0/Z$, where $J_0$ is
the flux of the incoming wave, and $D$ is the diffusion constant.
Thus
\begin{equation}
\langle I(z) \rangle= J_{0} \frac{Z-z}{Z}. \label{AvIDiff}
\end{equation}
This solution implies that the flux inside the sample is $\langle
J_z\rangle =-D \partial \langle I(z) \rangle /\partial z=J_0 \ell_{tr}/Z$
and therefore the average transmission coefficient through the slab
is ratio of the
 mean free path to the width of the slab:
\begin{equation}
 \langle T \rangle=\frac{\langle J_{z}\rangle}{J_{0}}=\frac{\ell}{Z}
\end{equation}
Notice that in our conventions the diffusion constant, $D$ has
dimensions of length, and is the transport mean free
path.

Consider now the density correlation function
(\ref{def:IntensityCo}). Solving
Eqs.~(\ref{eq:incompress}), (\ref{eq:J}), and calculating ${\cal C}
(r)$, using the correlation function of the Langevin sources
(\ref{eq:JLangevin}) (evaluated with the help of (\ref{AvIDiff})),
we obtain
\begin{equation}
{\cal C} ({\bf r}) = \langle  I(z) \rangle^2 \left\{ \begin{array}{ll}
\frac{1}{2k^2 r^2}   & \lambda \ll r\ll \ell_{tr} \\ \ & \  \\
\frac{3}{2 k^2\ell_{tr} r}    & \ell_{tr} \ll r \ll Z
\end{array} \right.,
\end{equation}
where it is assumed that the observation points are far form the end
of the slab, i.e. $\ell_{tr}\ll z \ll Z$.

The above result shows a power
law decay of the speckle correlations which is similar
to the case of directed waves. Yet, unlike directed wave, the
transmission coefficient of the system in diffusive systems
experience sample specific fluctuations. This is due to the finite
amount of backscattering which can be safely neglected in the case
of directed waves. In order to evaluate the magnitude of these
fluctuations, let us consider the integrated flux passing through
the slab:
 \begin{eqnarray}
\delta \bar{J}_z=\frac{1}{Z}\int_{V}d^3 r J_{z},
\end{eqnarray}
where $V$ denotes the volume of the slab. Here we assume the slab to
be finite with  dimensions $X$, $Y$ and $Z$, such that $Z \ll X,Y$.
Now, as follows from Eq.~(\ref{eq:J}), the current $J_{z}$ contains
two contributions:
\begin{eqnarray}
J_z= - D \frac{\partial}{\partial z} \delta I +J^{L}_z.  \label{eq:Jz}
\end{eqnarray}
The first contribution vanishes upon integration over space,
therefore the fluctuations in the total current are essentially due
to the contribution from the Langevin sources:
\begin{eqnarray}
\langle \left(\delta \bar{J}_z\right)^2 \rangle=
\frac{1}{Z^2}\int_{V}d^3 r d^3 r' \langle  J_{z}^L({\bf r})  J_{z}^L({\bf r}')\rangle.
\end{eqnarray}
Substituting Eq.~(\ref{eq:JLangevin}) for the correlation function of
the Langevin sources, and evaluating the integral we obtain:
\begin{eqnarray}
\langle \left(\delta \bar{J}_z\right)^2 \rangle= V \frac{\lambda^2 J_0
  \ell}{18 \pi}.
\end{eqnarray}
From here we conclude that the fluctuations in the transmission
coefficient scale as:
\begin{eqnarray}
\frac{\langle \delta T^2 \rangle}{\langle T \rangle^2} =
\frac{ \langle \left(\delta \bar{J}_z\right)^2 \rangle}{\langle
  J_z\rangle^2 V^2}= \frac{ \lambda^2 Z}{18 \pi \ell X Y}\propto \frac{1}{N},
\end{eqnarray}
where  $N\sim \nu V/\tau_{0} \sim \ell XY/\lambda^{2}Z$ is the total 
number of eigenfrequencies lying within frequency  band of 
width $1/\tau_{0}$, centered at the  frequency  of the incoming beam. 
Here $\nu \sim 1/ c \lambda^2$ is the density of states of the slab 
(per unit volume), $\tau_{0}=Z^{2}/\ell c$, is the 
typical time of diffusion through the sample, and $c$ is the wave velocity.

\subsubsection{Sensitivities of the speckle pattern in the diffusive regime}

Below we summarize the results of the speckle pattern sensitivities
to various perturbations in the diffusive regime. These results are
obtained by solving Eqs.~(\ref{sendif1}-\ref{sendif3}) and
identifying the the relevant scale of the perturbation parameter.

The sensitivity  to a change in the wave frequency is characterized
by the frequency scale of the order of
\begin{equation}
\omega^{*} = \frac{c \ell}{Z^{2}},
\end{equation}
where $c$ is the wave velocity, and $\ell$ is the elastic mean free
path. This frequency scale is the inverse time which takes the wave
to propagate through the sample.

The sensitivity to a change in the angle of the incoming wave,
$\phi$, is characterized by the scale
\begin{equation}
\phi^*= \frac{1}{k\sqrt{\ell Z}}. \label{phi*dif}
\end{equation}
The interpretation of this result is similar to that presented for
directed waves. Here, however, the diffusive nature of the ray
dynamics implies that the the original distance between two
 rays which reach the same point at the screen is of order
of $\sqrt{\ell Z}$. Therefore the interference of these rays
will become completely different when the phase difference, due to the
change in the incidence angle, is of order one, i.e. $k \sqrt{\ell Z} \phi^* \sim 1$. This condition leads to
(\ref{phi*dif}).

\begin{figure}
\includegraphics[width=8.0cm]{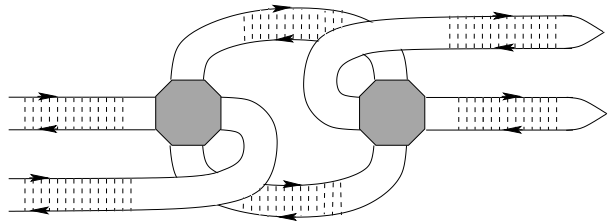}
\caption{The diagram contributing to the transmission coefficient
  correlations at large angles. The gray boxes are Hikami boxes,
  while solid lines connected by dashed lines represent averaged
  Green functions, see appendix A for details}
\label{fig:TwoHikamiBox}
\end{figure}

Finally let us discuss the sensitivity of the transmission
coefficient to a change in the angle of incidence, in a finite three
dimensional system. This sensitivity
may be described by the correlation function of the fluctuations
$\delta T(\boldsymbol{\theta})$ at two different angles, and the
result takes the form\cite{ZyuzinSpivakRev}
\begin{equation}
\frac{\langle \delta T(\boldsymbol{\theta})
\delta T(\boldsymbol{\theta'})\rangle}
{ \langle \delta T^{2}\rangle } \sim \left \{
\begin{array}{ll}
 \frac{3\lambda}{4\pi Z}
 \frac{1}{\left|\boldsymbol{\theta -\theta'}\right|} &
\textrm{~~if~~~
$\frac{\lambda}{Z}< |\boldsymbol{\theta -\theta'}|< \frac{\lambda}{\ell}$}, \\
\frac{\lambda^2}{Z\ell} &   \textrm{~~if~~~
$\frac{\lambda}{l}<|\boldsymbol{\theta-\theta'}|$}.
\end{array} \right. ~ \label{Tsen}
\end{equation}
As we show above, the fluctuations in the transmission coefficient
follow from the fluctuations in the current due to the Langevin
current sources. Therefore, one expects that the above correlation
function can be deduced from the correlation function of the
Langevin sources (\ref{sendif2}) where $\gamma$ stands for the
change in the incidence angle of the incoming wave.  This procedure,
indeed, gives the result within the range $\frac{\lambda}{Z}<
|\boldsymbol{\theta -\theta'}|< \frac{\lambda}{\ell}$. However for
larger difference in the angle of incidence, i.e.
$\frac{\lambda}{l}<|\boldsymbol{\theta-\theta'}|$, the behavior is
dominated by an additional contribution which is not described by
the Boltzmann-Langevin approach. This contribution can be calculated
from a diagram which contains two Hikami boxes, as shown
in Fig.~\ref{fig:TwoHikamiBox}. In real space it may be
associated with pair of orbits which intersect twice during their
 propagation in the system.

At this point it is instructive to mention the relation between
Eq.~(\ref{Tsen}) and the universal conductance fluctuations of
mesoscopic metals. The conductance in these systems is proportional
to the integral of the transmission coefficient over the angle,
$G\propto \int T(\boldsymbol{\theta}) d \boldsymbol{\theta}$.
Therefore according to Eq.~(\ref{Tsen}) the main contribution
to the conductance fluctuations,
\begin{equation}
\langle (\delta G)^{2}\rangle\propto \int d \boldsymbol{\theta} d
\boldsymbol{\theta'} \delta \langle T(\boldsymbol{\theta})\delta
T(\boldsymbol{\theta}')\rangle,
\end{equation}
comes from the interval of large angle difference, $\frac{\lambda}{l}<|
\boldsymbol{\theta}-\boldsymbol{\theta'}|$. As a result we have
$(\delta G)^{2}\rangle\sim e^{4}/\hbar^{2}$ for a three dimensional 
system where all dimensions are of the same order\cite{Altshuler,LeeStone}.

\section{Conclusions}\label{sec:conclusions}

We have developed a method of description of speckle statistics in
elastically scattering media which can be applied to both diffusive
and to the ballistic regime. Our main result is given by
Eqs.~(\ref{av-ray-dist}-\ref{eq:langevinsources}), which have a form
of kinetic equations with random sources. Though the derivation of
these equations in Appendix \ref{sec:Appendix_a} involved the Born
approximation for the amplitude of scattering on individual
scatterers, we believe that the region of the applicability of these
equation is much broader: They are valid as long as the Boltzmann
kinetic equation Eq.~(\ref{av-ray-dist}) holds. Namely, $\ell\gg
\lambda,\xi$, and $|{\bf r-r'}|\gg \lambda,\xi$.

We would like to mention that the results presented above
substantially differ from those known in the literature (see for
example Refs.~\onlinecite{Tatarski,Kravtsov,Prokhorov,Dashen}).
First, the correlation function (\ref{eq:main}) exhibits a universal
long range power law behavior over a wide range of distance, $\rho$.
The only  non-universal regimes are at the tail, $\rho\gg Z
\theta^2/\theta_{0}$ , and the short distance region, $\rho \sim
\xi$. This result is in contrast with the results presented in
Refs.~\onlinecite{Tatarski,Kravtsov,Prokhorov,Dashen} where the
intensity correlator ${\cal C}(\rho)$ depends on  the detailed form
of the disorder correlation function, $g(r)$, and usually
decays exponentially at $\rho>\xi$. Second,
${\cal C}(\rho)$ changes its sign
as a function of $\rho$. This property is a consequence of the current
conservation and it is absent from previous studies.
For instance, the sign change of ${\cal
C}(\rho)$ implies, that the fluctuations of the integrated intensity
over disks of radius $R>Z\theta$ is proportional to $R$, see
Eq.~(\ref{eq:power-asymptotic}), rather than $R^2$, as would follow
from Refs.~\onlinecite{Tatarski,Kravtsov,Prokhorov,Dashen}. 
These differences will affect interpretations of any measurement of
speckles done with the help of a sensor  aperture that is much larger
than the wavelength.

Finally we would like to mention that
our results  may be easily extended to cases with light
polarization, optically active media, Faraday effect, and coherent
short wave pulses as long as their duration is longer than
$\tau=\ell/c$. These issues are left for future studies.

This work has been supported by the Packard Foundation, by the NSF
under Contracts No. DMR-0228104, and by the Israel Science
Foundation (ISF) funded by the Israeli Academy of Science and
Humanities, and by the USA-Israel Binational Science Foundation
(BSF).

\appendix

\section{Derivation of the main equations}
\label{sec:Appendix_a}

The derivation of Eqs. (\ref{av-ray-dist}), (\ref{eq:langevin}), and
(\ref{eq:langevinsources}) is based on the standard impurity diagram
technique~\cite{Abrikosov}. And the relevant diagram blocks were
derived in numerous works. However in most cases the calculations
were done either for the case of delta-correlated disorder potential
or in the diffusive regime. In this paper we deal with a general
situation of an arbitrary angular dependence of the scattering
cross-section. Therefore below we outline the derivation of our
formalism and present expressions for the main diagram blocks.

The wave equation (\ref{eq:St-wave}) can be written in the form of a
stationary Schrodinger equation for a particle moving in the
presence of a random impurity potential,
\begin{equation}\label{eq:V_n}
      V(\boldsymbol{r}) = -2k^2\delta n(\boldsymbol{r}).
\end{equation}
The solution of Eq.~(\ref{eq:St-wave}) can be written as $\psi({\bf
r})= \int d {\bf r}' G^R({\bf r},{\bf r}') J({\bf r}')$, where
$J({\bf r}')$ is the source of radiation and $G^{R/A}({\bf r},{\bf
r}')$ is the retarded Green function, $ G^{R/A}\equiv \left(
k^2+\nabla^2-\hat{V} \pm i\eta
    \right)^{-1}$.
Here $\hat{V}$ denotes the impurity potential operator. This reduces
the problem of speckle statistics of coherent waves to that of
averaging products of retarded and advanced Green functions. The
latter problem can be treated using the impurity diagram
technique~\cite{Abrikosov}. We derive the expression for the various
diagram blocks below.

\subsection{Average Green function}
\label{sec:GF}

\begin{figure}
\includegraphics[width=7.0cm]{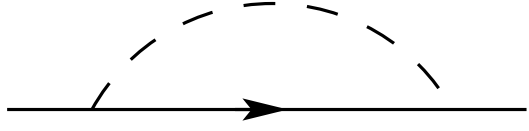}
\caption{The self-energy diagram in the Born approximation. }
\label{fig:self_en}
\end{figure}

In the Born approximation\footnote{In the literature on wave
propagation in disordered media the Born approximation
for the self energy is frequently referred to
as the Burret approximation~\cite{Tatarski}.} the self-energy,
$\Sigma(k,\boldsymbol{p})$, is given by a single diagram in
Fig.~\ref{fig:self_en}.

Its evaluation gives for the disorder-averaged Green function,
\begin{eqnarray}\label{eq:GF_wave}
    G^{R/A}(k,\boldsymbol{p})&=&
    \frac{1}{k^2-p^2\pm ik\ell^{-1}},
\end{eqnarray}
where the mean free path is given by Eqs.~(\ref{eq:l_def}) and
(\ref{kernel-wave}).

\subsection{Derivation of the Boltzmann equation}
\label{sec:appendix_Botzmann}

To derive the Boltzmann equation we will need to evaluate products
of Green functions at two different frequencies corresponding to
wave numbers, $ k_\pm=k\pm \delta k/2$.

 The spatial evolution of the
ray distribution function, Eq.~(\ref{Eq:raydis}), can be obtained by
expressing the solution of the wave equation in terms of the Green's
functions and performing disorder averaging. In the leading
approximation in $\lambda/\ell$ the ray distribution function
evolution is described by the sum of ladder diagrams.

Each disorder-averaged Green function is strongly peaked in the
momentum region where the on-shell condition is satisfied, $k=| {\bf
p} |$. In the limit of dilute scatterers the width of this peak,
$\sim 1/\ell$, is much smaller than the typical momentum momentum
transfer at each collision. Therefore the integration over the
magnitude of momenta in the Green functions can be carried out
separately and before the direction integration. Defining
$\mathbf{s}$ as the unit vector along the momentum $\mathbf{p}=p
\mathbf{s} $ we evaluate the product of the disorder-averaged
retarded and advanced Green's functions integrated over $p$,
\begin{eqnarray}\label{eq:cal_B}
   \frac{1}{ \mathcal{B}_{\delta k , \boldsymbol{q}}
    ( \boldsymbol{s})}
    &\equiv& 4\pi \int_0^\infty
    \frac{ p^2 dp}{2\pi^2} \,
G^R(k_+, p\boldsymbol{s}+\boldsymbol{q}/2)
    G^A(k_-,p\boldsymbol{s}-\boldsymbol{q}/2)=\frac{1}{-i\delta k +i
    \boldsymbol{s}\boldsymbol{q} +\ell^{-1}}.
\end{eqnarray}

The ray distribution function $f(\mathbf{s},\mathbf{q})$ is given by
the sum of ladder diagrams and can be expressed in a compact way
using the operator notations,
\begin{equation}\label{eq:f_oper}
    f=\sum_{n=0}^\infty \left(\hat{\cal B}^{-1} \hat{W} \right)^n
f_0  =\left(\hat{\cal B}- \hat{W}\right)^{-1}\hat{\cal B}f_0  .
\end{equation}

Here $f_0$ is the initial ray distribution function, $\hat{\cal B}$
is the integral operator whose kernel in the Fourier representation
is given by Eq.~(\ref{eq:cal_B}), and $\hat{W}$ is the integral
operator acting in the space of directions,
\begin{equation}\label{eq:hat_W_def}
    \hat{W}f(\boldsymbol{s})\equiv \int
    d\boldsymbol{s}'
    W(\boldsymbol{s}-\boldsymbol{s}')f(\boldsymbol{s}'),
    \quad W(\boldsymbol{s}-\boldsymbol{s}')\equiv
    \frac{1}{4\pi}w(k[\boldsymbol{s}-\boldsymbol{s}']).
\end{equation}

Multiplying Eq.~(\ref{eq:f_oper}) by $\left(\hat{\cal B}-
\hat{W}\right)$ from the left and using Eq.~(\ref{eq:l_def}) we
obtain the Boltzmann-Langevin equation,
\begin{eqnarray}\label{eq:Boltzmann_f}
    &&\left(-i\delta k + \boldsymbol{s} \cdot \nabla_r \right) f(\boldsymbol{s},\boldsymbol{r})
    - I_{st}[ f ] = \mathcal{L} , \\
    &&\mathcal{L}=\left(-i \delta k+\boldsymbol{s}\cdot
    \nabla_{{\bf r}} +
    \ell^{-1} \right) f^0(\boldsymbol{s},\boldsymbol{r}),
    \label{eq:L_f_0}
\end{eqnarray}
where the collision integral $I_{st}[f]$ is defined in
Eq.~(\ref{av-ray-dist}).

If one is interested in the average ray distribution function then
$f_0$ in Eq.~(\ref{eq:Boltzmann_f}) should be understood as the ray
distribution function of the incident radiation at the boundary of
the disordered medium. In this case the source vanishes in the
interior of the medium and the average ray distribution function
satisfies the usual homogeneous Boltzmann equation.

\subsection{Hikami box}

\begin{figure}
\includegraphics[width=7.0cm]{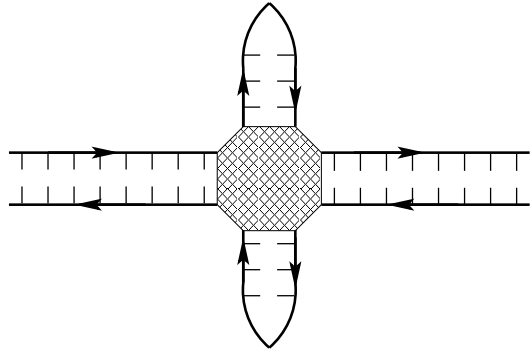}
\caption{The diagram for the irreducible correlator of ray distribution functions
at different points. Two impurity ladders emanating at the radiation source enter
the Hikami box, represented by the hashed octagon, from left and right. The
ladders going to the observation points leave the  Hikami box from the top and
the bottom.} \label{fig:correlator}
\end{figure}

Next let us consider the diagram in Fig.~\ref{fig:correlator} that represents the
irreducible correlator of the ray distribution functions. It allows the following
interpretation which is at the heart of the Boltzmann-Langevin approach developed
in this paper. The impurity ladders connecting the observation points to the
Hikami box propagate the fluctuations of the distribution function from the
Hikami box out to the observation points. This propagation is described by the
inhomogeneous Boltzmann-Langevin equation (\ref{eq:Boltzmann_f}). Then the right
hand side of Eq.~(\ref{eq:Boltzmann_f}) may be interpreted as the ``Langevin
force'' that results in the fluctuations of the ray distribution function. The
fluctuations of the Langevin force are described by the Hikami box connected to
the impurity ladders going out to the radiation sources. Since the latter define
the average ray distribution function we see that by evaluating the Hikami box we
will relate the fluctuations of the Langevin force to the average ray
distribution function.

\begin{figure}
\includegraphics[width=16.0cm]{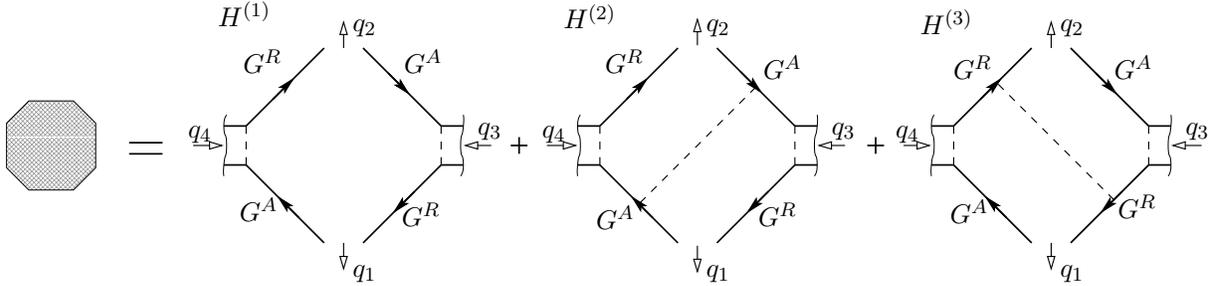}
\caption{The three diagrams for the Hikami box; $H^{(1)}$, $H^{(2)}$, and
$H^{(3)}$. The ladders coming from the radiation sources enter the Hikami box
from left and right and are characterized by the four-momenta
$q_{3/4}=(\omega_{3/4},\bm{q}_{3/4})$ and the ray directions $\bm{s}_{3/4}$. The
ladders going to the observation points exit the Hikami box from the top and the
bottom and are characterized by the four-momenta
$q_{1/2}=(\omega_{1/2},\bm{q}_{1/2})$ and the ray directions $\bm{s}_{1/2}$. Note
that in our notations the Hikami box contains a single impurity line for each of
the incoming ladders and no impurity lines for the outgoing ladders.}
\label{fig:Hikami}
\end{figure}

The Hikami box with the external legs is given by the three diagrams in
Fig.~\ref{fig:Hikami}. It is characterized by the four-momenta
$q_{i}=(\omega_{i},\boldsymbol{q}_{i})$ and the unit vectors $\boldsymbol{s}_i$
characterizing the ray directions. Here $i=1,2$ correspond to outgoing momenta
(ladders going to the observation points) and $i=3,4$ to the incoming ones
(ladders coming from the radiation source). The momenta satisfy the conservation
law, $q_1+q_2=q_3+q_4$.  The analytic expression that corresponds to the first
diagram (with no impurity line) is
\begin{eqnarray}\label{eq:H1}
    H^{(1)}_{ k}( \{ q_i\},\{\boldsymbol{s}_i \})&=&
    \delta(\boldsymbol{s}_1-\boldsymbol{s}_2)(4\pi)^2
    W(\boldsymbol{s}_1-\boldsymbol{s}_3)W(\boldsymbol{s}_2-\boldsymbol{s}_4)
    \int \frac{p^2 d p}{2\pi^2}
    G^R (k+\omega_1,\boldsymbol{p}+\boldsymbol{q}_1)
    G^A (k,\boldsymbol{p}) \nonumber \\
    && \times G^R (k+\omega_4,\boldsymbol{p}+\boldsymbol{q}_4)
    G^A (k + \omega_1-\omega_3,\boldsymbol{p} +\boldsymbol{q}_1 -\boldsymbol{q}_3)
    \nonumber \\
    &=& \frac{\pi}{k^2}\delta(\boldsymbol{s}_1-\boldsymbol{s}_2)
    \frac{W(\boldsymbol{s}_1-\boldsymbol{s}_3)W(\boldsymbol{s}_2-\boldsymbol{s}_4)
    }{\mathcal{B}_1\mathcal{B}_2}\left(
    \frac{1}{\mathcal{B}_3} +\frac{1}{\mathcal{B}_4}\right),
\end{eqnarray}
where we used the shorthand notation
$\mathcal{B}_i=\mathcal{B}_{\delta k_i , \boldsymbol{q}_i}
(\boldsymbol{s} )$ (with
$\boldsymbol{s}=\boldsymbol{s}_1=\boldsymbol{s}_2$) and utilized the
momentum conservation,
$\mathcal{B}_1+\mathcal{B}_2=\mathcal{B}_3+\mathcal{B}_4$.

The second diagram in Fig.~\ref{fig:Hikami} contains an impurity line connecting
the two advanced Green functions (between $q_3$ and $q_2$, and $q_4$ and $q_1$
respectively). It is given by the expression,
\begin{eqnarray}\label{eq:H2}
    H^{(2)}_{ k}( \{ q_i\},\{\boldsymbol{s}_i \})&=&
    (4\pi)^3
    W(\boldsymbol{s}_1-\boldsymbol{s}_3)W(\boldsymbol{s}_2-\boldsymbol{s}_4)
    W(\boldsymbol{s}_1-\boldsymbol{s}_2)
    \int \frac{p^2 d p}{2\pi^2}
    G^R (k+\omega_1,\boldsymbol{p}+\boldsymbol{q}_1)
    G^A (k,\boldsymbol{p}) \nonumber \\
    && \times
    G^A (k + \omega_1-\omega_3,\boldsymbol{p} +\boldsymbol{q}_1 -\boldsymbol{q}_3)
      \int \frac{p'^2 d p'}{2\pi^2}G^A (k,\boldsymbol{p}')
    G^R (k+\omega_4,\boldsymbol{p}'+\boldsymbol{q}_4) \nonumber \\
    &&\times G^A (k + \omega_1-\omega_3,\boldsymbol{p}' +\boldsymbol{q}_1
    -\boldsymbol{q}_3)\nonumber \\
    &=& -\frac{\pi}{k^2}
    \frac{W(\boldsymbol{s}_1-\boldsymbol{s}_3)W(\boldsymbol{s}_2-\boldsymbol{s}_4)
    W(\boldsymbol{s}_1-\boldsymbol{s}_2)}{\mathcal{B}_1\mathcal{B}'_2
    \mathcal{B}_3\mathcal{B}'_4},
\end{eqnarray}
where unprimed $\mathcal{B}$'s depend on $\boldsymbol{s}_1$ and
primed ones on $\boldsymbol{s}_2$,
\begin{eqnarray}
  \mathcal{B}_i&=&\mathcal{B}_{\omega_i , \boldsymbol{q}_i} ( \boldsymbol{s}_1), \\
  \mathcal{B}'_i&=&\mathcal{B}_{\omega_i ,
  \boldsymbol{q}_i} (\boldsymbol{s}_2 ).
\end{eqnarray}

The third diagram of the Hikami box contains an impurity line
connecting the two retarded Green functions (between $q_1$ and
$q_3$, and $q_2$ and $q_4$ respectively). It is given by
\begin{eqnarray}\label{eq:H3}
    H^{(3)}_{ k}( \{ q_i\},\{\boldsymbol{s}_i \})&=&
    (4\pi)^3
    W(\boldsymbol{s}_2-\boldsymbol{s}_3)W(\boldsymbol{s}_1-\boldsymbol{s}_4)
    W(\boldsymbol{s}_1-\boldsymbol{s}_2)
    \int \frac{p^2 d p}{2\pi^2}
    G^R (k+\omega_1,\boldsymbol{p}+\boldsymbol{q}_1)
    G^A (k,\boldsymbol{p}) \nonumber \\
    && \times
    G^R (k + \omega_4,\boldsymbol{p} +\boldsymbol{q}_4)
      \int \frac{p'^2 d p'}{2\pi^2} G^R (k+\omega_1,\boldsymbol{p}'+\boldsymbol{q}_1)
    G^R (k+\omega_4,\boldsymbol{p}'+\boldsymbol{q}_4) \nonumber \\
    &&\times G^A (k + \omega_1-\omega_3,\boldsymbol{p}' +\boldsymbol{q}_1
    -\boldsymbol{q}_3) \nonumber \\
    &=& -\frac{\pi}{k^2}
    \frac{W(\boldsymbol{s}_2-\boldsymbol{s}_3)W(\boldsymbol{s}_1-\boldsymbol{s}_4)
    W(\boldsymbol{s}_1-\boldsymbol{s}_2)}{\mathcal{B}_1\mathcal{B}'_2
    \mathcal{B}'_3\mathcal{B}_4}.
\end{eqnarray}

Next we make use of the fact that in Eqs.~(\ref{eq:H1}),
(\ref{eq:H2}), and (\ref{eq:H3}) the operators with indices $3$ and
$4$ act on impurity ladders $3$ and $4$ that go out to the radiation
sources. These ladders are equal to the average ray distribution
functions, $\langle f_{\boldsymbol{s}} \rangle$. In the interior of
the medium the latter obey the Boltzmann equation
(\ref{eq:Boltzmann_f}) with the vanishing right hand side, see
discussion below Eq.~(\ref{eq:L_f_0}). Therefore we have
\begin{equation}\label{eq:Boltzmann_f_W}
    \mathcal{B}^{-1}\hat{W}\langle f_{\boldsymbol{s}} \rangle=\langle
    f_{\boldsymbol{s}} \rangle .
\end{equation}
Using Eq.~(\ref{eq:Boltzmann_f_W}) and combining Eqs.~(\ref{eq:H1}),
(\ref{eq:H2}), and (\ref{eq:H3}) we obtain the correlator of the
Langevin forces that enter the right hand side of
Eq.~(\ref{eq:Boltzmann_f}). As a result we can describe speckle
fluctuations in the framework of the Boltzmann-Langevin scheme,
Eqs.~(\ref{eq:langevin}) and (\ref{eq:langevinsources}).

\section{Derivation of formula (\ref{eq:main})}
\label{sec:appendix_B}

In this appendix we derive formula (\ref{eq:main}) for the intensity
correlation function in the directed waves limit. For this purpose
we employ the parabolic and the Markov approximations. Namely the
scalar wave equation (\ref{eq:St-wave}) is approximated by a simpler
equation, obtained by substituting $\psi \to e^{ikz} \psi({\bf r})$
into (\ref{eq:St-wave}) and neglecting second order derivatives of
the wave function with respect to $z$. The resulting equation takes
the form of a Schrodinger equation where the coordinated associated
with the propagation direction, $z$, plays the role of fictitious
time:
\begin{equation}
 i\frac{\partial \psi}{\partial z}=-\frac{1}{2k}
 \left(\frac{\partial^2}{\partial x^2}+
 \frac{\partial^2}{\partial y^2} \right)\psi+ k \delta n({\bf r})\psi,
\label{eq:parabolic}
\end{equation}
The analysis of this equation is further simplified when the Markov
approximation is employed. The latter corresponds to the situation
where the disorder correlation function is anisotropic: It is delta
correlated in the propagation direction $z$, and long ranged
correlated in the perpendicular directions:
\begin{equation}
\langle \delta n({\bf  r}) \delta n({\bf r}') \rangle
=g_\perp({\mathbf \rho}-{\mathbf \rho}' )\delta(z-z'). \label{markov}
\end{equation}
Here angular brackets denote disorder averaging, and
$g_\perp(\boldsymbol{\rho})$ represents the disorder correlation
function in the $\boldsymbol{\rho}=(x,y)$ space. We shall assume
that $\delta n({\bf  r})$ is gaussian random function and that
$g_\perp(\boldsymbol{\rho})$ is isotropic.

These approximations however, do not imply, necessarily, diffusive
motion, and therefore applies also for length scales shorter than
the mean free path. Within these approximations the Green function
associated with, Eq. (\ref{av-ray-dist}), henceforth called
``diffuson'' and denoted by ${\cal D}(\boldsymbol{ \rho},{\bf
p};z)$, satisfies an equation of the form:
\begin{eqnarray}
\left( \frac{ \partial }{\partial z} +
\frac{{\bf p}}{k} \frac{ \partial  }{\partial \boldsymbol{  \rho}}
 \right) {\cal D}(\boldsymbol{ \rho},{\bf p};z)
- k^2 \int \frac{ d^2 q}{4\pi^2}  \hat{g}_\perp({\bf q})
\left( {\cal D}(\boldsymbol{ \rho},{\bf p}-{\bf q};z)-
{\cal D}(\boldsymbol{ \rho},{\bf p};z) \right)=\delta (\boldsymbol{
\rho}-\boldsymbol{ \rho}_0) \delta ({\bf p}-{\bf p}_0) \delta(z),
\end{eqnarray}
where $\hat{g}_\perp({\bf q})$ is the Fourier transform of
$g_\perp(\rho)$. Notice that here the momentum ${\bf p}$
is a two component vector in the space perpendicular to the
propagation direction.

The above equation can be simplified by Fourier transforming it
with respect to the momentum, ${\bf p}$. Thus
if denote by ${\bf x}$ the  variable conjugate to the
momentum ${\bf p}$, and $\hat{\cal D}$ denotes the Fourier transform of the
diffuson ${\cal D}$, then
\begin{eqnarray}
\left( \frac{ \partial }{\partial z} -
\frac{i}{k} \frac{ \partial^2}{\partial \boldsymbol{  \rho}\partial
  {\bf x}}
 \right) \hat{\cal D} - k^2 \left[ g_\perp(x)-g_\perp(0) \right]
\hat{\cal D}= \frac{e^{i{\bf p}_0 {\bf x}}}{4\pi^2} \delta (\boldsymbol{
\rho}-\boldsymbol{ \rho}_0) \delta(z).
\end{eqnarray}
To solve this equation we further take its Fourier transform with
respect to $z$ and  $\boldsymbol{\rho}$ (with conjugate variables
denoted by $q_z$, and ${\bf q}$ respectively):
\begin{equation}
\left( i q_z  +
\frac{\boldsymbol{q}}{k} \frac{ \partial}{\partial {\bf x}}
 \right) \hat{\cal D} - k^2 \left[ g_\perp(x)-g_\perp(0) \right]
\hat{\cal D}= \frac{e^{i{\bf p}_0 {\bf x}-i\boldsymbol{ q}
  \boldsymbol{ \rho}_0}}{4\pi^2}.
\end{equation}
Now let us decompose the vector ${\bf x}$ into its components:
$x_\parallel$ parallel to the vector  $\boldsymbol{q}$, and
 $x_\perp$ perpendicular to that vector. Then the solution of the
above equation takes the form:
\begin{equation}
\hat{\cal D}=\frac{k e^{-i\boldsymbol{q} \boldsymbol{
      \rho}_0}}{4\pi^2 q}
      \int_{-\infty}^{x_\parallel}dx_\parallel'e^{i
 p_{0\parallel} x_\parallel'+ip_{0\perp} x_\perp}\exp
      \frac{k}{q} \left[
      \int_{x_\parallel'}^{x_\parallel} dx_\parallel''
\left( iq_z- k^2 \left( g_\perp
      (x_\parallel'',x_\perp)-g_\perp (0) \right)\right) \right],
 \end{equation}
where under the assumption of isotropy in the plane perpendicular to
the propagation direction $g_\perp (x_\parallel,x_\perp) =g_\perp
\left( \sqrt{x^2_\parallel +x_\perp^2}\right)$. Now, taking the
inverse Fourier transform with respect to $q_z$, integrating over
$x_\parallel'$, and Fourier transforming the result with respect to
${\bf x}$ we obtain the result for the diffuson:
\begin{equation}
{\cal D}(\boldsymbol{ \rho},{\bf p};\boldsymbol{ \rho}_0,{\bf p}_0;z) =
\int \frac{d^2 x}{4\pi^2} \int \frac{d^2q}{4 \pi^2} e^{i\boldsymbol{
  q}(\boldsymbol{ \rho}- \boldsymbol{ \rho}_0)+i {\bf p}_0\left( {\bf x}- \frac{z}{k}\boldsymbol{ q} \right)-i{\bf p} {\bf x}} \exp \left[
      \frac{k^3}{q}
 \int_{x_\parallel-\frac{q}{k} z}^{x_\parallel} dx_\parallel''
\left(  g_\perp
      (x_\parallel'',x_\perp)-g_\perp (0) \right) \right].
\label{ballistic-diffuson}
\end{equation}
If we assume boundary conditions where the
average distribution function, at $z=0$, is given by
$\bar{f}_0(\boldsymbol{ \rho}, {\bf p})$, then
for $z>0$ the average distribution function is given by the integral:
\begin{equation}
\bar{f}(\boldsymbol{ \rho},{\bf p},z)= \int d^2 \rho_0 d^2 p_0 {\cal
  D}(\boldsymbol{ \rho},{\bf p};\boldsymbol{ \rho}_0,{\bf p}_0;z)
\bar{f}_0(\boldsymbol{ \rho}_0, {\bf p}_0)
\end{equation}
In particular assuming the incident wave, at
$z=0$, to be a plane wave pointing at the $z$ direction,
$\bar{f}_0(\boldsymbol{ \rho}, {\bf p})= 4 \pi^2 I_0 \delta ({\bf p})$ where
$I_0$ is the density, the above integral reduces to
\begin{equation}
\bar{f}({\bf p},z)= I_0 \int d^2 x
\exp \left[-i {\bf p}{\bf x}- 2k^2 \left(  g_\perp(x)-g_\perp (0)
\right) \right] \label{exactfbar}
\end{equation}
This formula is exact assuming the parabolic and the Markov
approximation.  Namely it holds as long as  $l \gg \xi$ (Markov
approximation), and $\xi \gg \lambda$ (small angle scattering, i.e.
parabolic approximation). It holds for any distance $z<l_{tr}$, and
for any value of the momentum ${\bf p}$. It may be further
simplified if we assume $z \gg \ell$ where $\ell$ is the elastic
mean free path. In this case the dynamics is of diffusive nature in
the
 angle of directions and one may approximate the correlation
function $g_\perp(x)$ using Taylor expansion near $x=0$:
\begin{equation}
g_\perp(x)\simeq g_\perp(0)- D_\theta x^2\label{g-expand}
\end{equation}
where $D_\theta= -g_\perp''(0)/2$ ($g_\perp''(x)$ denote the second
derivative of $g_\perp(x)$ with respect to $x$) is the angular
diffusion constant. Substituting (\ref{g-expand}) into
(\ref{exactfbar}) and preforming the integral over $x$ yields:
\begin{equation}
\bar{f}({\bf p},z)\simeq \frac{\pi I_0}{k^2 D_\theta z}
\exp \left[-\frac{p^2}{4 k^2 D_\theta z} \right], ~~~~~z\gg l
\label{diff-approx-av}
\end{equation}

Let us now consider the fluctuations of the distribution
function. Using Eqs.~(\ref{eq:langevin}) and
(\ref{eq:langevinsources}), one may write their corresponding
correlation function  as
\begin{eqnarray}
\langle \delta f(\boldsymbol{ \rho},{\bf p},Z)\delta f(\boldsymbol{ \rho}',{\bf
  p}',Z) \rangle= \int_0^Z dz \int d^2 \rho'' d^2p_1 d^2p_2
{\cal D}(\boldsymbol{ \rho},{\bf p}; \boldsymbol{ \rho}'',{\bf
  p}_1;Z-z)~~~~~~~~~~~~~~~~~~~~~~~~~~~~~~~~~~\label{correlation-int}
\\  \times
\left( g_\perp(0)\delta({\bf p}_1-{\bf p}_2)- \hat{g}_\perp({\bf p}_1-{\bf p}_2)
\right) \bar{f}({\bf p}_1,z)\bar{f}({\bf p}_2,z)
{\cal D}(\boldsymbol{ \rho}',{\bf p}'; \boldsymbol{ \rho}'',{\bf p}_2;Z-z) \nonumber
\end{eqnarray}
where, as before, this result has been obtained under the parabolic and the
Markov approximations. The density correlation function, ${\cal
  C}(\boldsymbol{\rho-\rho'}$ can be deduced from
(\ref{correlation-int}) by integration over ${\bf p}$ and ${\bf  p}'$:
\begin{equation}
C(\boldsymbol{ \rho}-\boldsymbol{ \rho}') =\int d^2p d^2p' \langle \delta f(\boldsymbol{ \rho},{\bf p},Z)\delta f(\boldsymbol{ \rho}',{\bf
  p}',Z) \rangle
\end{equation}
Thus substituting (\ref{correlation-int}) and
(\ref{ballistic-diffuson}) into the above formula, a and
performing the integral over $ \boldsymbol{ \rho}''$ yields
\begin{eqnarray}
C(\boldsymbol{ \rho})=
\int_0^Z d\zeta \int \frac{d^2 q}{4 \pi^2}\frac{d^2 p_1}{4 \pi^2}
\frac{d^2 p_2}{4 \pi^2}  \left(  g_\perp(0)\delta({\bf p}_1-{\bf
    p}_2)- \hat{g}_\perp({\bf p}_1-{\bf p}_2)
\right) \bar{f}({\bf p}_1,\zeta)\bar{f}({\bf p}_2,\zeta)
~~~~~~\label{correlation-N-int}
\\  \times
 \exp \left[ i \boldsymbol{ q} \left(
    \boldsymbol{ \rho} - \frac{{\bf p}_1-{\bf
    p_2}}{k}(Z-\zeta)\right)+ 2k^2 \int_0^{Z-\zeta} d\eta \left( g(\eta q/k)-g(0)
    \right) \right]
 \nonumber
\end{eqnarray}
This formula is obtained essentially by introducing one Hikami box
into the diagrams. The small parameter controlling this
approximation (i.e. the neglect of additional Hikami boxes) is $l
\gg \xi^2/\lambda$. The distance between the observation points should
be larger than the disorder correlation length, $|\boldsymbol{
\rho}-\boldsymbol{ \rho}'|> \xi$.

The above integral can be further simplified if we assume the width
of the system, $Z$, to be much larger than the elastic mean free
path, $Z\gg l$. In that case, as can be seen from formula
(\ref{diff-approx-av}), the width of the average distribution
function $\bar{f}({\bf p},\zeta)$ at $\zeta \gg \ell$  is much wider
than the width of $\hat{g}_\perp({\bf p})$, as the width of first
function is $k \sqrt{D_\theta \zeta} \sim
\frac{1}{\xi}\sqrt{\zeta/\ell}$, while the second  is of order
$1/\xi$. Therefore, assuming the integral over $z$ to be dominated by
points near the screen $z=Z$ (an assumption which turns out to be
consistent) one may approximate the factor $\bar{f}({\bf
  p}_1,\zeta)\bar{f}({\bf p}_2,\zeta)$ in the integral (\ref{correlation-int})
as  $\bar{f}({\bf
  p}_1,\zeta)\bar{f}({\bf p}_2,\zeta) \simeq \bar{f}^2({\bf p}_1,\zeta))$, and
consider ${\bf p}_1$ and $\tilde{\bf p}={\bf p}_2-{\bf p}_1$ as
independent variables.  Since in this regime $\bar{f}({\bf p},\zeta)$ is
given by Eq.~(\ref{diff-approx-av}) the integral over $p_1$ and
$p_2-p_1$ can be performed and the result takes is
\begin{eqnarray}
C(\rho)=
\frac{\pi I_0^2}{D_\theta} \int_\ell^Z \frac{d\zeta}{\zeta}
\int \frac{d^2 q}{4 \pi^2}  \left( g_\perp(0)- g_\perp\left( \frac{q
(Z-\zeta)}{k}\right) \right) \exp \left[ i \boldsymbol{ q}
    \boldsymbol{ \rho}+ 2k^2 \int_0^{Z-\zeta}
d\eta \left( g(\eta q/k)-g(0)    \right) \right].
\end{eqnarray}
Performing the angular part of the integral over $\boldsymbol{ q}$,
expressing the pre-exponential factor as a derivative of the
exponent, and changing the integration variable from $\zeta$ to
$Z-\zeta$ we finally obtain formula (\ref{eq:main}):
\begin{eqnarray}
C(\rho)=
\frac{ I_0^2}{4 D_\theta k^2} \int_0^{Z-\ell} \frac{d\zeta}{\zeta-Z}
\int q d q J_0(q \rho) \frac{d}{d\zeta} \exp \left[ -\frac{2}{\ell}
  \int_0^{\zeta} d\eta \left(1- \tilde{g}\left(\frac{q}{k} \eta\right)
\right) \right],
\end{eqnarray}
where $\ell^{-1}= k^2
g_\perp(0)$, while $\tilde{g}(\eta)= g_\perp(\eta)/g_\perp(0)$.


\begin{thebibliography}{99}
\bibitem{clav} The procedure of averaging in the classical
 case, and the concept of the fluctuations of the distribution 
function $\delta \tilde{f}$ require some clarification.
The classical kinetic equation Eq.~(\ref{kinav}) is valid in systems
which exhibit chaotic partiocle dynamics, which means that the
kinetics of the system has a self-averaging character. Thus the
averaging procedure denoted by the double brackets $\langle\langle
\ldots \rangle\rangle$ does not include averaging over the random
configurations of the scattering potential. The classical kinetic
scheme does not necesserily require an introduction of the
fluctuations of the distribution function $\delta \tilde{f}$. There
is another approach to claculation of classical nonequilibrium
fluctuations \cite{GurevichGanzevichKatilus}, which is equvalent to
Eqs.~(\ref{kinav}), (\ref{Classical-langevin}), and (\ref{kinSK}),
and is based on equations for the sigle $\langle \langle
\tilde{f}\rangle\rangle$ and two particle $\langle \langle \delta
\tilde{f}\delta \tilde{f}\rangle\rangle$ distribution functions.
\end{thebibliography}
\end{document}